\documentclass[twocolumn,amsmath,nofootinbib]{revtex4} 
\usepackage{url}
\usepackage{amssymb}
\usepackage{amsfonts}
\usepackage{amsbsy}
\usepackage{graphicx}
\usepackage{hyperref}
\usepackage{array} 
\usepackage{multirow}

\newcolumntype{x}[1]{%
>{\centering\hspace{0pt}}p{#1}}%
\providecommand{\openone}{\leavevmode\hbox{\small1\kern-3.8pt\normalsize1}}

\def\ie{{\frenchspacing\it i.e.}}
\def\eg{{\frenchspacing\it e.g.}}
\def\etc{{\frenchspacing\it etc.}}

\def\spose#1{\hbox to 0pt{#1\hss}}
\def\simlt{\mathrel{\spose{\lower 3pt\hbox{$\mathchar"218$}}
   \raise 2.0pt\hbox{$\mathchar"13C$}}}
\def\simgt{\mathrel{\spose{\lower 3pt\hbox{$\mathchar"218$}}
     \raise 2.0pt\hbox{$\mathchar"13E$}}}
 \def\simpropto{\mathrel{\spose{\lower 3pt\hbox{$\mathchar"218$}}
     \raise 2.0pt\hbox{$\propto$}}}

\def\beq#1{\begin{equation}\label{#1}}
\def\eeq{\end{equation}}
\def\beqa#1{\begin{eqnarray}\label{#1}}
\def\eeqa{\end{eqnarray}}
\def\eq#1{equation~(\ref{#1})}	
\def\Eq#1{Equation~(\ref{#1})}
\def\eqn#1{~(\ref{#1})}

\def\fig#1{Figure~\ref{#1}}
\def\Fig#1{Figure~\ref{#1}}

\def\Tab#1{Table~\ref{#1}}

\def\Sec#1{Section~\ref{#1}}

\def\ed{\end{document}}

\def\p{{\bf p}}
\def\q{{\bf q}}
\def\rvec{{\bf r}}

\def\x{{\bf x}}

\def\xb{\bar{\x}}
\def\xh{\widehat{\x}}

\def\I{{\bf I}}
\def\A{{\bf A}}
\def\Ab{\bar{\bf A}}
\def\Ap{\hbox{\bf\'{A}}}
\def\Ah{\widehat{\bf A}}
\def\At{\tilde{{\bf A}}}
\def\B{{\bf B}}
\def\C{{\bf C}}

\def\Ch{\widehat{{\bf C}}}

\def\D{{\bf D}}

\def\m{{\bf m}}
\def\n{{\bf n}}
\def\nb{\bar{\bf n}}
\def\nh{\widehat{\bf n}}
\def\M{{\bf M}}
\def\N{{\bf N}}

\def\SS{{\mathbf\Sigma}}
\def\SSb{\bar{\mathbf\Sigma}}
\def\SSh{\widehat{\mathbf\Sigma}}
\def\SSt{\tilde{\mathbf\Sigma}}

\def\T{{\bf T}}

\def\Y{{\bf Y}}

\def\tr{\hbox{tr}\,}

\def\nth{n^{\rm th}}

\def\tensormult{\otimes}
\def\expec#1{\langle#1\rangle}

\def\rn{}
\def\nn#1 #2{#2. #1}				
\def\nnn#1 #2 #3{#2. #3. #1}			
\def\nnnn#1 #2 #3 #4{#2. #3. #4 #1}		
\def\nnnnn#1 #2 #3 #4 #5{#2. #3. #4 #5. #1}	

\def\rf#1;#2;#3;#4;#5 {{\frenchspacing\par\rn#1, #3 {\bf #4}, #5 (#2). \par}}
\def\rfe#1;#2;#3;#4 {{\frenchspacing\par\rn#1, #3, #4 (#2). \par}}
\def\rg#1;#2;#3;#4;#5;#6 {{\frenchspacing\par\rn#1, #3 {\bf #4}, #5 (#2). \par}}
\def\rfbook#1;#2;#3;#4;#5 {{\frenchspacing\par\rn#1, {\it #3} (#5, #4, #2).\par}}
\def\rfprep#1;#2;#3 {{\par\frenchspacing\rn#1, #3 (#2).\par}}
\def\rfproc#1;#2;#3;#4;#5;#6 {{\frenchspacing\par\rn#1 #2, in {\it #3}, ed. #4 (#5: #6)\par}}
\def\rfprocp#1;#2;#3;#4;#5;#6;#7 {{\frenchspacing\par\rn#1 #2, in {\it #3}, ed. #4 (#5: #6), p#7\par}}

\begin{document}
\pdfoptionalwaysusepdfpagebox=5


\title{Improved Measures of Integrated Information}

\author{Max Tegmark}

\address{Dept.~of Physics \& MIT Kavli Institute, Massachusetts Institute of Technology, Cambridge, MA 02139}
\address{Theiss Research, La Jolla, CA 92037}

\date{November 21 2016, published in {\it PLoS Comput.~Biol.}~{\bf 12} (11): e1005123, doi:10.1371/journal.pcbi.1005123}


\vspace{10mm}

\begin{abstract}
Although there is growing interest in measuring integrated information in computational and cognitive systems, current methods for doing so in practice are computationally unfeasible.
Existing and novel integration measures are investigated and classified by various desirable properties. 
A simple taxonomy of $\Phi$-measures is presented where they are each characterized by their choice of factorization method (5 options),  choice of probability distributions to compare ($3\times 4$ options) and choice of measure for comparing probability distributions (7 options). When requiring the $\Phi$-measures to satisfy a minimum of attractive properties, these hundreds of options reduce to a mere handful, some of which turn out to be identical.
Useful exact and approximate formulas are derived that can be applied to real-world data from laboratory experiments without posing unreasonable computational demands.
\end{abstract} 
  
\maketitle

\section{Introduction}
\label{IntroSec}



What makes an information-processing system conscious in the sense of having a subjective experience?
Although many scientists used to view this topic as beyond the reach of science, the study of Neural Correlates of Consciousness (NCCs) has become quite mainstream in the neuroscience community in recent years --- see, \eg,  \cite{rees2002neural,metzinger2000neural}.
To move beyond correlation to causation \cite{chalmers1995facing}, neuroscientists have begun searching for a theory of consciousness that can predict what physical phenomena cause consciousness (defined as subjective experience \cite{chalmers1995facing}) to occur. Dehaene \cite{dehaene2014toward} reviews a number of candidate theories currently under active discussion, including the {\it Nonlinear Ignition} model (NI) \cite{dehaene2011conscious,shadlen2011consciousness}, 
the {\it Global Neuronal Workspace} (GNW) model 
\cite{dehaene2001towards,shanahan2005applying,dehaene1998neuronal}
and {\it Integrated Information Theory} (IIT) \cite{tononi2008consciousness,oizumi2014phenomenology}.
Rapid progress in artificial intelligence is further fueling interest in such theories and how they can be generalized  
to apply not only to biological systems, but also to engineered systems such as computers and robots and ultimately arbitrary arrangements of elementary particles \cite{tegmark2014consciousness}.

Although there is still no consensus on necessary and sufficient conditions for a physical system to be conscious, there is broad agreement that it needs to be able to store and process information in a way that is somehow {\it integrated}, 
not consisting of nearly independent parts. As emphasized by Tononi \cite{tononi2008consciousness}, it must be impossible to decompose a conscious system into nearly independent parts --- otherwise these parts would feel like two separate conscious entities. 
While integration as a {\it necessary} condition for consciousness is rather uncontroversial, IIT goes further and makes the bold and controversial claim that it is also a {\it sufficient} condition for consciousness, using an elaborate mathematical integration definition \cite{oizumi2014phenomenology}.

\begin{table*}
{\footnotesize
\begin{tabular}{|l|l|ccccccccccccccc|}
\hline
&											&$\phi^{2.5}$&$\phi^{2.5'}$&$\phi^{2.5''}$&$\phi^{3.0}$&$\phi^M$&$\phi^B$&$\phi^{MD}$&$\phi^M_{kk'}$&$\phi^{oak}$&$\phi^{opk}$&$\phi^{ots}$&$\phi^{ofu}$&$\phi^{nas}$&$\phi^{mas}$&$\phi^{xfk}$\\
\hline
\parbox[t]{2mm}{\multirow{4}{*}{\rotatebox[origin=c]{90}{Major}}}
&Always {\bf non-negative}						&  y&  y&  y&  y&  y&\N&  y&  y&  y&  y&  y&  y&  y&  y&  y\\
&Always {\bf finite} even for $\infty$-dimensional system	&\N&  y&  y&\N&  y&  y&  y&  y&  y&  y&  y&  y&\N&  y&  y\\
&Vanishes for {\bf deterministic} system	 (drawback)	&  n&  n&  n&  n&  n&  n&  n&  n&  n&  n&  n&  n&  n&  n&\Y\\
&Vanishes for {\bf separable} system					&  y&  y&  y&  y&  y&\N&  y&  y&  y&  y&\N&  y&  y&  y&  y\\
\hline
\parbox[t]{2mm}{\multirow{4}{*}{\rotatebox[origin=c]{90}{Minor}}}
&Vanishes for {\bf afferent} system					&  y&  y&  y&  y&\N&\N&\N&\N&  y&\N&\N&\N&  y&  y&\N\\
&Vanishes for {\bf efferent} system					&  y&  y&  y&  y&\N&\N&\N&\N&\N&  y&\N&\N&\N&\N&\N\\
&{\bf State-dependent}							&  y&  y&  y&  y&\N&\N&\N& y&  y&  y&\N&\N&\N&\N& y\\
&Based on {\bf symmetric} probability distance			&\N&\N&\N&  y&\N&\N&\N&\N&\N&\N&\N&\N&\N&\N&\N\\
\hline
&Intuitively {\bf interpretatable}						&2&2&2&2&2&0&2&2&2&2&0&1&0&0&0\\
&Computationally {\bf tractable}						&1&2&2&0&2&2&2&2&2&2&2&2&1&2&2\\
\hline
\end{tabular}
\caption{Properties of different integration measures. All but the third are desirable properties;
capitalized N/Y (no/yes) indicate when an integration measure lacks a desirable property or has an undesirable one.
The first four properties are generally agreed to be important, while the 
second set of four have been argued to be important by some authors. Interpretability refers to the extent to which the measure can be given an 
information-theoretic interpretation satisfying desirable properties of integration (see text). Computability refers to the feasibility of evaluating 
the measure in practice (see text).
\label{PropertyTable}
}
}
\end{table*}

\begin{table*}
{\footnotesize
\renewcommand{\arraystretch}{2.0}
\begin{tabular}{|l|l|l|}
\hline
Name&		Definition&Formula for Gaussian variables\\
\hline
$\phi^{\rm otu}\>(\phi^M)$	&$I(\x^A,\x^B)-I(\x_0^A,\x_0^B)$			&${1\over 2}\log {|\T_A| |\T_B| |\C|\over |\T| |\C_A| |\C_B|}
={1\over 2}\log{|\widehat\SS_A| |\widehat{\SS}_B|\over |\SS|}
$\\
$\phi^B$			&$I(\x_0,\x_1) - I(\x^A_0,\x^A_1) - I(\x^B_0,\x^B_1)$	&${1\over 2}\log {|\C|^2 |\T_A| |\T_B|\over |\T| |\C_A|^2|\C_B|^2}
={1\over 2}\log{|\C||\SSh_A|\SSh_B|\over |\SS||\C_A||\C_B|}$\\
$\phi^{otum}$ $(\phi^{MD})$	&$d_{MD}(p,q), \quad q_{ii'jj'} = {p_{ii'\cdot\cdot} p_{i\cdot j\cdot} p_{\cdot i'\cdot j'}\over p_{i\cdot\cdot\cdot}p_{\cdot i'\cdot\cdot}}$&
${1\over 2}\left[\ln{|\Ch|\over|\SSh|}
+\tr\left(\Ch^{-1}\C-\SSh^{-1}\SS-\Ap^t\SSh^{-1}\Ap\C\right)\right]$ for $\beta=1$,  $\Ch\equiv\Ah\C\Ah^t+\SSh$\\
$\phi^{\rm ots}$	&$I(\x^A,\x^B)$							&${1\over 2}\log {|\T_A| |\T_B|\over |\T|}$\\
$\phi^{\rm ofs}$	&$I(\x_1^A,\x_1^B)$	 						&${1\over 2}\log {|\C_A| |\C_B|\over |\C|}$\\
$\phi^{\rm ofu}$	&$-\sum\limits_{jj'} p_{\cdot\cdot jj'}\log\sum\limits_{ii'}{p_{i\cdot j\cdot}p_{\cdot i'\cdot j'}p_{ii'\cdot\cdot}\over p_{i\cdot\cdot\cdot}p_{\cdot i'\cdot\cdot}p_{\cdot\cdot jj'}}$	&${1\over 2}\left[\ln{|\C_q|\over|\C|}+\tr \C_q^{-1}\C-n\right]$,\quad$\C_q\equiv\Ah\C\Ah^t+\SSh$\\

$\phi^{\rm ofk}_{kk'}\>(\phi^M_{kk'})$	&$\sum\limits_{jj'}{p_{kk'jj'}\over p_{kk'\cdot\cdot}}\log{p_{kk'jj'}p_{k\cdot\cdot\cdot}p_{\cdot k'\cdot\cdot}\over p_{kk'\cdot\cdot}p_{k'\cdot j\cdot}p_{\cdot k'\cdot j'}}$
	&${1\over 2}\left[\x_0^t(\Ah-\A){\SSh}^{-1}(\Ah-\A)\x+\ln{|\SS_A|\,|\SS_B|\over|\SS|}+\tr{\SSh}^{-1}\SS-n\right]$\\
$\phi^{\rm oak}_{kk'}$		&$\sum\limits_j{p_{kk'j\cdot}\over p_{kk'\cdot\cdot}}\log{p_{kk'j\cdot}p_{k\cdot\cdot\cdot}\over p_{kk'\cdot\cdot}p_{k\cdot j\cdot}}$	&${1\over 2}\left[\Delta\m^t\SSb_A^{-1}\Delta\m+\ln{|\SSb_A|\over|\SS_A|}+\tr \SSb_A^{-1}\SS_A-n_A\right]$,\quad$\Delta\m\equiv(\A_A-\Ah_A)\x_0^A+\A_B\x_0^B$\\
$\phi^{\rm opk}_{kk'}$	&$\sum\limits_i{p_{i\cdot kk'}\over p_{\cdot\cdot kk'}}\log{p_{i\cdot kk'}p_{\cdot\cdot k\cdot}\over p_{\cdot\cdot kk'}p_{i\cdot k\cdot}}$&${1\over 2}\left[\Delta\m^t\tilde{\SSb}_A^{-1}\Delta\m+\ln{|\tilde{\SSb}_A|\over|\tilde{\SS}_A|}+\tr\tilde{\SSb}_A^{-1}\tilde{\SS}_A-n_A\right]$,\quad$\Delta\m\equiv(\At_A-\hat{\tilde{\A}}_A)\x_0^A+\At_B\x_1^B$\\

\hline



$\phi^{\rm xfk}_{kk'}$	&$I(\x_1^A,\x_1^B|\x_0=kk')$				&${1\over 2}\log {|\SS_A| |\SS_B| \over |\SS|}$\\


\hline
$\phi^{\rm nas}$	&$-\sum\limits_j p_{\cdot\cdot j\cdot}\log\sum\limits_{ii'}{p_{ii'j\cdot}p_{i\cdot\cdot\cdot}\over n_B p_{ii'\cdot\cdot}p_{\cdot\cdot j\cdot}}$&$\infty$\\
$\phi^{\rm nak}_k$	&$-\sum\limits_j {p_{k\cdot j\cdot}\over p_{k\cdot\cdot\cdot}} \log\sum\limits_{i'}{p_{ki'j\cdot}p_{k\cdot\cdot\cdot}\over n_B p_{ki'\cdot\cdot}p_{k\cdot j\cdot}}$&$\infty$\\

$\phi^{\rm nps}$	&$-\sum\limits_i p_{i\cdot\cdot\cdot}\log\sum\limits_{jj'}{p_{i\cdot jj'}p_{\cdot\cdot j\cdot}\over n_B p_{\cdot\cdot jj'}p_{i\cdot\cdot\cdot}}$&$\infty$\\
$\phi^{\rm npk}_k$	($\phi^{\rm 2.0}_k$) &$-\sum\limits_i {p_{i\cdot k\cdot}\over p_{\cdot\cdot k\cdot}} \log\sum\limits_{j'}{p_{i\cdot kj'}p_{\cdot\cdot k\cdot}\over n_B p_{\cdot\cdot kj'}p_{i\cdot k\cdot}}$&$\infty$\\

\hline
$\phi^{\rm mas}$	&$-\sum\limits_j p_{\cdot\cdot j\cdot}\log\sum\limits_{ii'}{p_{ii'j\cdot}p_{i\cdot\cdot\cdot}p_{\cdot i'\cdot\cdot}\over p_{ii'\cdot\cdot}p_{\cdot\cdot j\cdot}}$	&${1\over 2}\left[\ln{|\C_q|\over|\C_A|}+\tr \C_q^{-1}\C_A-n_A\right]$,\quad$\C_q\equiv\SS_A+\A_A\C_A\A_A^t+\A_{AB}\C_B\A_{AB}^t$\\
$\phi^{\rm mak}_k$	&$-\sum\limits_j {p_{k\cdot j\cdot}\over p_{k\cdot\cdot\cdot}} \log\sum\limits_{i'}{p_{ki'j\cdot}p_{k\cdot\cdot\cdot}p_{\cdot i'\cdot\cdot}\over p_{ki'\cdot\cdot}p_{k\cdot j\cdot}}$&${1\over 2}\left[\Delta\m^t\SSb_A^{-1}\Delta\m+\ln{|\SSb_A|\over|\SS_A|}+\tr \SSb_A^{-1}\SS_A-n_A\right]$,\quad$\Delta\m\equiv(\Ah_A-\A_A)\x_0^A$\\
$\phi^{\rm mps}$	&$-\sum\limits_i p_{i\cdot\cdot\cdot}\log\sum\limits_{jj'}{p_{i\cdot jj'}p_{\cdot\cdot j\cdot}p_{\cdot\cdot\cdot j'}\over p_{\cdot\cdot jj'}p_{i\cdot\cdot\cdot}}$&${1\over 2}\left[\ln{|\C_q|\over|\C_A|}+\tr \C_q^{-1}\C_A-n_A\right]$,\quad$\C_q\equiv\SSt_A+\At_A\C_A\At_A^t+\At_{AB}\C_B\At_{AB}^t$\\
$\phi^{\rm mpk}_k$	&$-\sum\limits_i {p_{i\cdot k\cdot}\over p_{\cdot\cdot k\cdot}} \log\sum\limits_{j'}{p_{i\cdot kj'}p_{\cdot\cdot k\cdot}p_{\cdot\cdot\cdot j'}\over p_{\cdot\cdot kj'}p_{i\cdot k\cdot}}$	&${1\over 2}\left[\Delta\m^t\tilde{\SSb}_A^{-1}\Delta\m+\ln{|\tilde{\SSb}_A|\over|\SSt_A|}+\tr \tilde{\SSb}_A^{-1}\SSt_A-n_A\right]$,\quad$\Delta\m\equiv(\hat{\At}_A-\At_A)\x_1^A$\\
%
%
\hline
$\phi^{2.5}$			&$\min\{\phi^{\rm nak},\phi^{\rm npk}\}$&$\infty$\\
$\phi^{2.5'}$			&$\min\{\phi^{\rm mak},\phi^{\rm mpk}\}$&\\
$\phi^{2.5''}$			&$\min\{\phi^{\rm oak},\phi^{\rm opk}\}$&\\
\hline
\end{tabular}
\caption{Integration $\phi$ for different measures.
$\A\equiv\B^t\C^{-1}$, $\A_b\equiv\B\C^{-1}=\SS\A^t\SS_b^{-1}$,
$\SS\equiv\C-\B^t\C^{-1}\B=\C-\A\C\A^t$ and
$\SS_b\equiv\C-\B\C^{-1}\B^t=\C-\A_b\C\A_b^t= [\C^{-1}+\A^t\SS^{-1}\A]^{-1}=\C-\C\A^t\C^{-1}\A\C$. $\C$ is the data covariance matrix and $\B$ is the cross-covariance between different times as defined by \eq{Meq2}.
 }
\label{MeasureTable}
}
\end{table*}

As neuroscience data improves in quantity and quality, it is timely to resolve this controversy  
by testing the many experimental predictions that IIT makes \cite{oizumi2014phenomenology} with state-of-the-art laboratory measurements.
Unfortunately, such tests have been hampered by the fact that the integration measure proposed by IIT is computationally infeasible to evaluate for large systems, growing super-exponentially with the system's information content.
This has lead to the development of various alternative integration measures that are simpler to compute or have other desirable properties. For example, 
Barrett \& Seth \cite{barrett2011practical} proposed an attractive integration measure that is easier to compute from neuroscience data, but whose interpretation is complicated by the fact that it can be negative in some cases \cite{seth2011causal,oizumi2016measuring}. 
\cite{casali2013theoretically} used an integration measure inspired by complexity theory to successfully predict who was
conscious in a sample including patients who were awake, in deep sleep, dreaming, sedated and with 
locked-in syndrome.  \cite{sitt2014large} suggest that state transition entropy correlates with consciousness.
Griffith \& Koch have proposed defining integration of a system as the synergistic information that its parts have about the future, which appears promising although there does not yet exist a unique formula for it \cite{griffith2014quantifying}.
Even the team behind IIT has updated their integration measure twice through successive refinements of their theory \cite{tononi2008consciousness,oizumi2014phenomenology}.
Despite these definitional and computational challenges, interest in measuring integration is growing, not only in neuroscience but also in other fields, ranging from physics \cite{tegmark2014consciousness} and evolution \cite{edlund2011integrated} to the study of collective intelligence in social networks \cite{engel2015groups}.

It is therefore interesting and timely to do a comprehensive investigation of existing and novel integration measures, classifying them by various desirable properties. This is the goal of the present paper, as summarized in \Tab{PropertyTable} and \Tab{MeasureTable}.
The rest of this paper is organized as follows. 
In \Sec{MeasureSec}, we investigate general integration measures and their properties, presenting our results in \Sec{TaxonomySec}.
In \Sec{GaussianSec}, we derive useful formulas for many of these measures that can be applied to the sort of time-series data that is 
typically measured in laboratory experiments, with continuous variables.
We explore further algorithmic speedups and approximations in \Sec{ApproximationSec} and summarize our conclusions in \Sec{ConclusionsSec}.

\section{Measures of integration}
\label{MeasureSec}

Following Tononi \cite{tononi2008consciousness}, we will use the symbol $\Phi$ to denote integrated 
information.\footnote{Note that our analysis is focused {\it only} on integration, not on consciousness; 
besides integration, a true measure of consciousness may involve additional requirements that this paper does not consider.
For example, Scott Aaronson has criticized in a widely read blog post the claim that integration is a sufficient condition for consciousness, and 
IIT discusses postulates including cause-effect power, composition and exclusion \cite{oizumi2014phenomenology}.
}
All measures of $\Phi$ aim to quantify the extent to which a system is interconnected, 
yielding $\Phi=0$ if the system consists of two independent parts, and a larger $\Phi$ the more the parts affect each other.
Mathematically,  all $\Phi$-measures are defined in a two-step process:
\begin{enumerate}
\item Given an imaginary cut that partitions the system into two parts, define
a measure $\phi$ of how much these two parts affect each other.  \Tab{MeasureTable} lists many $\phi$-options.
\item Define $\Phi$ as the $\phi$-value for the ``cruelest cut" that minimizes $\phi$.
A major numerical challenge is that the number of cuts to be 
minimized over grows super-exponentially with the number of bits in the system.
A further challenge in this step is how to best handle cuts splitting the system into parts of unequal size.
\end{enumerate}

\begin{figure}[phbt]
\centerline{\includegraphics[width=88mm]{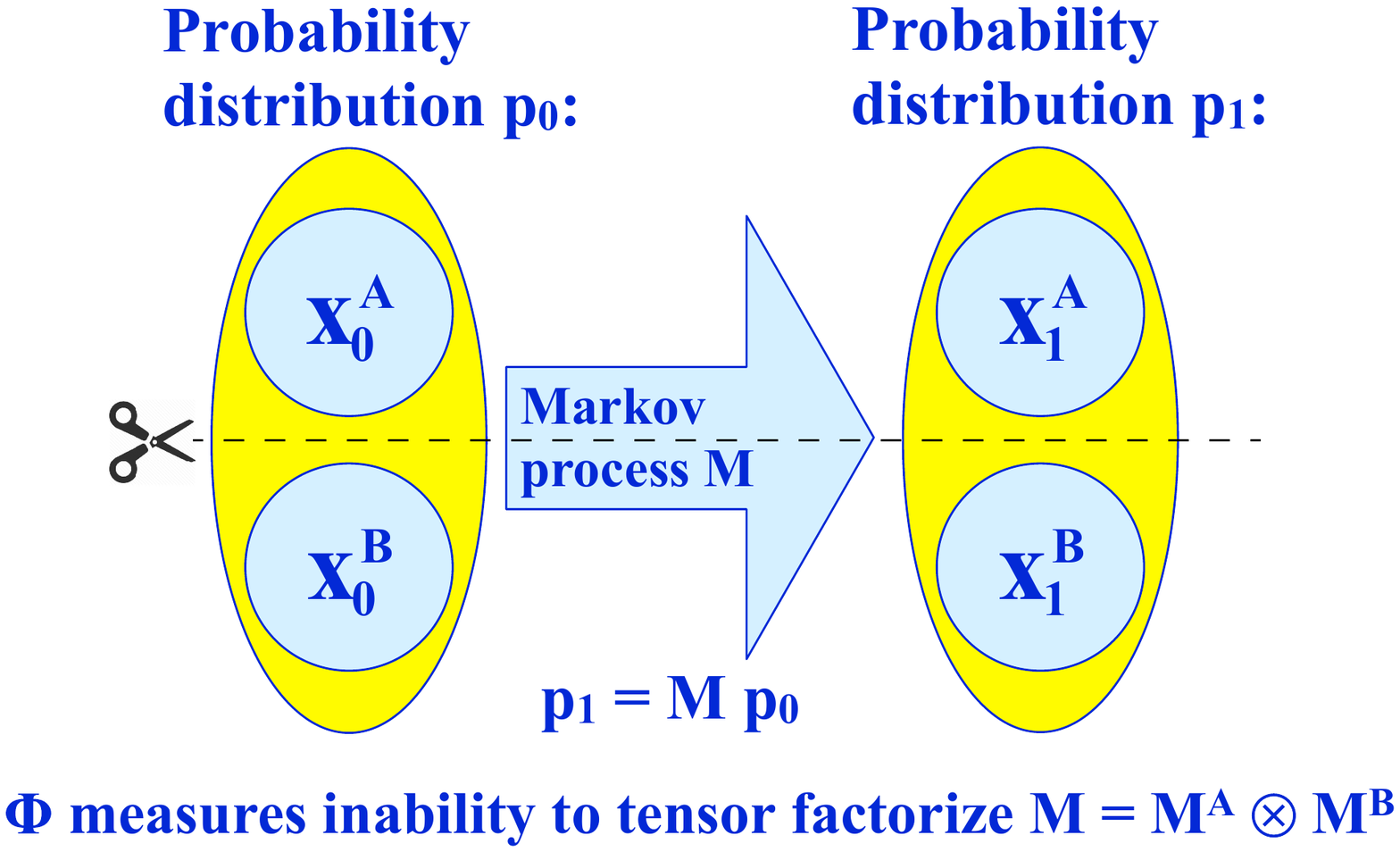}}
\caption{We model the time-evolution of the system state as a Markov process defined by a transition matrix $\M$:
when the (possibly unknown) system state evolves from $\x_0$ to $\x_1$, the corresponding probability distribution
evolves from $\p_0$ to $\p_1\equiv\M\p_0$.
All competing definitions of $\Phi$ quantify the inability to tensor factorize $\M$, which corresponds to approximating the system
as two disconnected parts A and B that do not affect one another.
\label{MarkovFig}
}
\end{figure}

Before delving into the many different options for defining $\Phi$, let us first introduce convenient notation 
general enough to describe all proposed integration measures, as illustrated in \fig{MarkovFig}.

\subsection{Interpreting evolution as a Markov process}

Consider two random vectors $\x_0$ and $\x_1$ whose joint probability distribution is 
$p(\x_0,\x_1).$ We will interpret them as the state of a time-dependent system $\x(t)$ at two separate times
$t_0$ and $t_1$.
For example, if these are two vectors of 5 bits each, then $p$ is a table of $2^{10}$ numbers giving the probability of each possible bit string, while if these are two vectors in 3D space, then $p$ is a function of 6 real continuous variables. 
We obtain the marginal distribution $p^{(n)}(\x_n)$ for the $\nth$ vector, where $n=0$ or $n=1$, by summing/integrating $p$ over the other vector. 

Below we will often find it convenient to denote these vectors as single indices $i=\x_0$ and $j=\x_1$. For example, this allows us to write the marginal distribution $p_0(\x_0)$
as $\sum_j p_{ij},$ 
where the sum over $j$ is to be interpreted as summation/integration over all allowed values of $\x_1$.
We also adopt the notation where replacing an index by a dot means that this index is to be summed/integrated over.
This lets us write the marginal distributions $p^{(0)}(\x_0)$ and $p^{(1)}(\x_1)$ as
\beq{MargDistEq}
p^{(0)}_i = p_{i\cdot}\quad\hbox{and}\quad p^{(1)}_j = p_{\cdot j}
\eeq

As illustrated in \fig{MarkovFig}, it is always possible to model this relation between $\x_0$ and $\x_1$ as resulting from a 
Markov process, where $\x_1$ is causally determined by a combination of $\x_0$ and random effects.
If we write the marginal distributions from \eq{MargDistEq}
as vectors $\p^{(0)}$ and $\p^{(1)}$, this Markov process is defined by
\beq{MarkovEq}
\p^{(1)} = \M\p^{(0)},
\eeq
where the Markov matrix $M_{ji}$ specifies the probability that a state $i$ transitions to a state $j$, 
and satisfies the conditions $M_{ji}\ge 0$ (non-negative transition probabilities)
and $M_{\cdot i}=1$ (unit column sums, guaranteeing probability conservation).
The standard rule for conditional probabilities gives
\beqa{ConditionalProbEq}
p_{ij}&=&P(\x_0=i\>\&\>\x_1=j) = \nonumber\\
&=&P(\x_0=i) P(\x_1=j|x_0=i) = p^{(0)}_i M_{ji},
\eeqa
which uniquely determines the Markov matrix as
\beq{MarkovMatrixEq}
M_{ji}={p_{ij}\over p^{(0)}_i} = {p_{ij}\over p_{i\cdot}},
\eeq
which is seen to satisfy the Markov requirements $M_{ji}\ge 0$ and $M_{\cdot i}=1$.

Note that any system obeying the laws of classical physics can be accurately modeled as a Markov process as long as the time step $\Delta t\equiv t_1-t_0$ is sufficiently short (defining $\x(t)$ as the position in phase space). If the process has ``memory" such that the next state depends not only on the current state but also on some finite number of past states, it can reformulated
as a standard memoryless Markov process by simply expanding the definition of the state $\x$ to include elements of the past.\footnote{Note that although full knowledge of the Markov 
matrix $\M$ completely specifies the dynamics of the system, a person wishing to compute its integration may not know $\M$ exactly. 
If $\M$ is not known from having built the system or having examined its inner workings, then passively observing it in action (without active interventions) 
may not provide enough information to fully reconstruct $\M$ \cite{chicharro2012two}. \Sec{GaussianSec} describes a convenient class of systems where $\M$ is relatively easy to determine.}

\subsection{A taxonomy of integration measures}


We will now see that this Markov process interpretation allows us create a simple taxonomy of integration measures $\phi$ that quantify the interaction between two subsystems. The idea is to approximate the Markov process by a {\it separable} Markov process that does not mix information between subsystems, and  to define the integration as a measure of how bad the best such approximation is.
Consider the system $\x$ as being composed of two subsystems $\x^A$ and $\x^B$, so that
the elements of the vector $\x$ are simply the union of the elements of $\x^A$ and $\x^B$, and let us define
the probability distribution
\beq{ptensorDefEq}
p_{ii'jj'}\equiv P(\x^A_0=i\>\&\>\x^B_0=i'\>\&\>\x^A_1=j\>\&\>\x^B_1=j').
\eeq
(For brevity, we will sometimes refer to this distribution $p_{ii'jj'}$ as simply $p$ below, suppressing the indices, and we will sometimes write $\x$ without indices
to refer to the full state at both times.)
The Markov matrix of \eq{MarkovMatrixEq} then takes the form
\beq{MarkovMatrixEq2}
M_{jj'ii'}={p_{ii'jj'}\over p_{ii'\cdot\cdot}}.
\eeq
The Markov process of \eq{MarkovEq} is separable if the Markov matrix $\M$ is a tensor product 
$\M^A\tensormult\M^B$, \ie, if
\beq{MarkovTensorEq}
M_{jj'ii'}= M^A_{ji}M^B_{j'i'}
\eeq
for Markov matrices $\M^A$ and $\M^B$ that determine the evolution of $\x^A$ and $\x^B$.

If our system is integrated so that $\M$ cannot be factored as in \eq{MarkovTensorEq}, we can nonetheless choose to approximate $\M$ by a matrix of the factorizable form 
$\M^A\tensormult\M^B$. If we retain the initial probability distribution $p_{ii'\cdot\cdot}$ for $\x_0$ but replace the correct Markov matrix $\M$ by the separable approximation $\M^A\tensormult\M^B$, then \eq{MarkovMatrixEq2} shows that the probability distribution
\beq{pDefEq}
 p_{ii'jj'}=M_{jj'ii'} \>p_{ii'\cdot\cdot}
 \eeq
 gets replaced by the probability distribution $q_{ii'jj'}$ given by
\beq{qdefEq}
q_{ii'jj'} = M^A_{ji}M^B_{j'i'} \>p_{ii'\cdot\cdot}
\eeq
which is an approximation of $p_{ii'jj'}$.
If $\M$ is factorizable (meaning that there is no integration), we can factor $\M$ such that the two probability distributions $q_{ii'jj'}$ and $p_{ii'jj'}$ are equal and, conversely, if 
the two probability distributions are different, we can use how different they are as an integration measure $\phi$.

To define an integration measure $\phi$ in this spirit, we thus need to make four different choices, which collectively specify it fully and determine where the $\phi$-measure belongs in our taxonomy:
\begin{enumerate}
\item Choose a recipe defining an approximate factorization $\M\approx\M^A\tensormult\M^B$.
\item Choose which probability distributions $p$ and $q$ to compare for exact and approximate $\M$ (the distribution for $\x$, $\x_1$ or $\x_1^A$, say).
\item Choose what to treat as known about 
$p_{ii'\cdot\cdot}$
when computing these probability distributions.
\item Choose a metric for how different the two probability distributions $p$ and $q$ are.
\end{enumerate}
These four options are described in Tables~\ref{FactorizationTable}, \ref{ComparisonTable} and \ref{MetricTable}, and we will now explore them in detail.

\begin{table*}
{\footnotesize
\renewcommand{\arraystretch}{1.8}
\begin{tabular}{|l|l|l|l|l|l|l|l|l|}
\hline
Code&Factorization method						&$M^A_{ji}$													&$M^B_{j'i'}$												&$\tilde{M}^A_{ij}$													&$\tilde{M}^B_{i'j'}$		&State-dependent?\\
\hline
n&Noising 									&${1\over n_B}\sum\limits_{i'}{p_{ii'j\cdot}\over p_{ii'\cdot\cdot}}$			&${1\over n_A}\sum\limits_i{p_{ii'\cdot j'}\over p_{ii'\cdot\cdot}}$		&${1\over n_B}\sum\limits_{j'}{p_{i\cdot jj'}\over p_{\cdot\cdot jj'}}$			&${1\over n_A}\sum\limits_j{p_{\cdot i'jj'}\over p_{\cdot\cdot jj'}}$		&N\\
m&Mild noising 								&$\sum\limits_{i'}{p_{ii'j\cdot}p_{\cdot i'\cdot\cdot}\over p_{ii'\cdot\cdot}}$	&$\sum\limits_i{p_{ii'\cdot j'}p_{i\cdot\cdot\cdot}\over p_{ii'\cdot\cdot}}$	&$\sum\limits_{j'}{p_{i\cdot jj'}p_{\cdot\cdot\cdot j'}\over p_{\cdot\cdot jj'}}$	&$\sum\limits_j{p_{\cdot i'jj'}p_{\cdot\cdot j\cdot}\over p_{\cdot\cdot jj'}}$	&N\\
o&Optimal not knowing state $\x_0$					&${p_{i\cdot j\cdot}\over p_{i\cdot\cdot\cdot}}$							&${p_{\cdot i'\cdot j'}\over p_{\cdot i'\cdot\cdot}}$					&${p_{i\cdot j\cdot}\over p_{\cdot\cdot j\cdot}}$							&${p_{\cdot i'\cdot j'}\over p_{\cdot\cdot\cdot j'}}$					&N\\
x&Optimal given $\x_0$  							&${p_{kk'j\cdot}\over p_{kk'\cdot\cdot}}$								&${p_{kk'\cdot j'}\over p_{kk'\cdot\cdot}}$							&${p_{kk'j\cdot}\over p_{kk'\cdot\cdot}}$								&${p_{kk'\cdot j'}\over p_{kk'\cdot\cdot}}$							&Y\\
a&Optimal given $\x_0$, on average					&${p_{i\cdot j\cdot}\over p_{i\cdot\cdot\cdot}}$							&${p_{\cdot i'\cdot j'}\over p_{\cdot i'\cdot\cdot}}$					&${p_{i\cdot j\cdot}\over p_{\cdot\cdot j\cdot}}$							&${p_{\cdot i'\cdot j'}\over p_{\cdot\cdot\cdot j'}}$					&N\\
\hline
\end{tabular}
\caption{Different options for approximate factorizations $\M\approx\M^A\tensormult\M^B$ and $\tilde{\M}\approx\tilde{\M}^A\tensormult\tilde{\M}^B$.
These options correspond to the first superscript in $\phi$-measures such as $\phi^{ofuk}$.
The optimal factorizations maximize the accuracy of the approximate probability distribution that they predict, 
while the ``noising'' factorizations are instead defined by treating the input from the other subsystem as random noise, either with uniform distribution (option ``n'') or with the
observed marginal distribution (option ``m'').
\label{FactorizationTable}
}
}
\end{table*}

\subsection{Options for approximately factoring $\M$}
\label{FactorizationSec}

\Tab{FactorizationTable} lists five factoring options which all have attractive features, and we will now describe each in turn.

\subsubsection{Approximately factoring $\M$ using noising}

The first option corresponds to the ``noising'' method used in IIT \cite{tononi2008consciousness}: 
the time evolution of one part of the system ($\x^A$, say) is determined from the past state $\x^A_0$ alone, 
treating $\x^B_0$ as random noise with some probability distribution $p^{(B0)}$ that is independent of $\x^A_0$.
In other words, we replace the initial probability distribution 
$p^{(0)}_{ii'}=p_{ii'\cdot\cdot}$ by the separable distribution
$p^{(0)}_{ii'}=p^{(A0)}_i p^{(B0)}_{i'}$.
We will now see that if we start with \eq{MarkovEq}, \ie, the Markov equation $\p^{(1)} = \M\p^{(0)}$, then this noising prescription gives
$\p^{(A1)} = \M^A\p^{(A0)}$ for a particular matrix $\M^A$.
 \Eq{MarkovEq} states that
\beq{MarkovEq2}
p^{(1)}_{jj'}=\sum_{ii'}M_{jj'ii'}p^{(0)}_{ii'}\>,
\eeq
and substituting the separable ``noising'' form of $p^{(0)}_{ii'}$ from above gives
\beq{NoisingDerivationEq}
p^{(A1)}_{j}\equiv p^{(1)}_{j\cdot}=\sum_{ii'}M_{j\cdot ii'}p^{(A0)}_i p^{(B0)}_{i'}
= \sum_iM^A_{ji} p^{(A0)}_{i}\>,
\eeq
where we have defined
\beq{NoisedMAeq}
M^A_{ji}\equiv\sum_{i'}M_{j\cdot ii'} p^{(B0)}_{i'}=\sum_{i'}{p_{ii'j\cdot}p^{(B0)}_{i'}\over p_{ii'\cdot\cdot}}.
\eeq
IIT chooses the noise to have maximum entropy, \ie, a uniform distribution over the $n_B$ possible states of subsystem B \cite{tononi2008consciousness}: 
\beq{IITnoisedMAeq}
p^{(B0)}_{i'} = {1\over n_B}.
\eeq
\Tab{FactorizationTable} lists the $\M^A$-matrix corresponding to this noising choice as well as the analogous $\M^B$-matrix.

\subsubsection{Approximately factoring $\M$ using mild noising}

One drawback of this choice is that uniform distributions are undefined for continuous variables such as measured voltages, because they cannot be normalized.
This means that any $\phi$-measure based on this noising factorization is undefined and useless for continuous systems.
This problem can be solved by adopting another natural choice for the noise distribution:
\beq{MildMAeq}
p^{(B0)}_{i'}=p_{\cdot i'\cdot\cdot},
\eeq
\ie, simply the marginal distribution for $\x^B_0$.
We term this option ``mild noising'', since the noise is less extreme (its entropy is lower) than with the previous noising option.
\Tab{FactorizationTable} lists the $\M^A$-matrix corresponding to this mild noising choice as well as the analogous $\M^B$-matrix.

\subsubsection{Optimally factoring $\M$}

A drawback of both factorizations that we have considered so far is that they might overestimate integration: there may exist an alternative factorization that is better in the sense of giving a smaller $\phi$.
The natural way to remedy this problem is to define $\phi$ by minimizing over all factorizations. 
This elegantly unifies with the fact that capital $\Phi$ is defined by minimizing over all partitions of the system into two parts:
we can capture both minimizations by simply saying ``minimize over all factorizations", since the choice of a tensor factorization includes a choice of partition.


In practice, the definition of the optimal factorization depends on what we optimize. 
We discuss various options below, and identify three particularly natural choices which are listed in \Tab{FactorizationTable}.
The first option makes the approximate probability distribution $q_{ii'jj'}$ as similar as possible to $p_{ii'jj'}$, where similarity is quantified by KL-divergence.
The second option treats the present state $\x_0$ as known and makes the conditional probability distribution for the future state $\x_1$ as similar as possible to the correct distribution.
This factorization thus depends on the state and hence on time, whereas all the others we have considered are state-independent. 
The third option is the factorization that minimizes this state-dependent $\phi$ {\it on average}; we will prove below that this factorization is identical to the first option.

In summary, \Tab{FactorizationTable} lists five factorization options that each have various attractive features; options 3 and 5 turn out to be identical. 
It is easy to show that if the Markov matrix $\M$ is factorizable (which means that the probability distribution is separable as $p_{ii'jj'}=p^A_{ij}p^B_{i'j'}$),
then all five factorizations coincide, all giving $M^A_{ji}=p^A_{ji}/p^A_{i\cdot}$ and $M^B_{j'i'}=p^A_{i'j'}/p^A_{i'\cdot}$.
This means that they will all agree on when $\phi=0$; otherwise the noising factorizations will yield higher $\phi$ than an optimized factorization.

\begin{table*}
{\footnotesize
\renewcommand{\arraystretch}{1.5}
\begin{tabular}{|l|l|l|l|l|l|l|}
\hline
&											&					&			&\multicolumn{3}{c|}{Conditioning option}\\
\cline{5-7}
	&										&					&			&u							&s								&k\\
	&										&					&			&$\x_t$ unknown				&$\x_t$-distribution separable				&$\x_t$ known\\
Code&Comparison option							&$p$				&$q$		&$q^u$ ($p^{(0)}_{ij}=p_{ii'\cdot\cdot}$)	&$q^s$ ($p^{(0)}_{ij}=p_{i\cdot\cdot\cdot}p_{\cdot i'\cdot\cdot}$)		&$q^k$ ($p^{(0)}_{ij}=\delta_{ik}\delta_{i'k'}$)\\
\hline
t	&Two-time state								&$p_{ii'jj'}$\			&$q_{ii'jj'}$	&$M^A_{ji}M^B_{j'i'}p^{(0)}_{ii'}$	&$\left(M^A_{ji}p^{(0)}_{i\cdot}\right)\left(M^B_{j'i'}p^{(0)}_{\cdot i'}\right)$	&$M^A_{ji}\delta_{ik}M^B_{j'i'}\delta_{i'k'}$\\
f	&Future state								&$p_{\cdot\cdot jj'}$		&$q_{\cdot\cdot jj'}$	&$\sum\limits_{ii'}M^A_{ji}M^B_{j'i'}p^{(0)}_{ii'}$		&$\left(\sum\limits_i M^A_{ji}p^{(0)}_{i\cdot}\right)\left(\sum\limits_{i'}M^B_{j'i'}p^{(0)}_{\cdot i'}\right)$	&$M^A_{jk}M^B_{j'k'}$\\
a	&Future state of subsystem A					&$p_{\cdot\cdot j\cdot}$	&$q_{\cdot\cdot j\cdot}$		&$\sum\limits_i M^A_{ji}p^{(0)}_{i\cdot}$	&$\sum\limits_i M^A_{ji}p^{(0)}_{i\cdot}$	&$M^A_{jk}$\\
p	&Past state of subsystem A						&$p_{i\cdot \cdot\cdot}$	&$\tilde{q}_{i\cdot \cdot\cdot}$	&$\sum\limits_j\tilde{M}^A_{ij}p^{(1)}_{j\cdot}$	&$\sum\limits_j\tilde{M}^A_{ij}p^{(1)}_{j\cdot}$	&$\tilde{M}^A_{ik}$\\
\hline
\end{tabular}
\caption{Different options for which probability distributions $p$ and $q$ to compare, 
corresponding to the second and third superscripts in $\phi$-measures such as $\phi^{ofuk}$.
The last three columns specify the formula for $q$ for the three conditioning options we consider: when the state $\x_0$ is unknown (u), has a separable probability distribution (s) and is known  (k), respectively.
\label{ComparisonTable}
}
}
\end{table*}

\subsection{Options for which probability distributions to compare}
\label{ComparisonSec}

\Tab{ComparisonTable} lists four options for which probability distributions $p$ and $q$ to compare. 
Arguably the most natural option is to simply compare the full distributions $p_{ii'jj'}$ and $q_{ii'jj'}$ that  describe our knowledge of the system at both times (the present state and the future state).
Another obvious option is to merely compare the predictions, \ie, the probability distributions 
$p_{\cdot\cdot jj'}$ and $q_{\cdot\cdot jj'}$ for the future state.
A third interesting option is to compare merely the predictions for one of the two subsystems (which we without loss of generality can take to be subsystem A), thus comparing $p_{\cdot\cdot j\cdot}$ and $q_{\cdot\cdot j\cdot}$.

Generally, the less we compare, the easier it is to get a low $\phi$-value. 
To see this, consider a system where $A$ affects $B$ but $B$ has no effect on $A$.
We could, for example, consider $A$ to be photoreceptor cells in your retina and $B$ to be the rest of your brain.
Then the second comparison option (``f") in \Tab{ComparisonTable} would give $\phi>0$ because we predict the future of your brain worse if we ignore the information flow from your retina, 
while the third comparison option (``a") in the table would give $\phi=0$ because the rest of your brain does not help predict the future of your retina.
In other words,  comparison option ``a'' makes $\phi$ vanish for {\it afferent} pathways, where information flows only inward toward the rest of the system.

IIT argues that any good $\phi$-measure indeed {\it should} vanish for afferent pathways, because a system can only be conscious if it can have effects on itself --- other systems that it is affected by without affecting will act merely as parts of its unconscious outside world \cite{tononi2008consciousness}.
Analogously, IIT argues that any good $\phi$-measure should vanish also for {\it efferent} pathways,
where information flows only outward away from the rest of the system.
The argument is that other systems that the conscious system affects without being affected by will again be unconscious, acting merely as unconscious parts of the outside world as far as the conscious system is concerned.

Option ``p'' in \Tab{ComparisonTable} has this property of $\phi$ vanishing for efferent pathways.
It is simply the time-reverse of option ``a'', 
quantifying the ability of $x^A_1$ to determine its past cause $x^A_0$ instead of 
quantifying the ability of $x^A_0$ to determine its future effect $x^A_1$.

To formalize this, consider that there is nothing in the probability distribution $p_{ii'jj'}$ that breaks time-reversal symmetry and says that we must interpret causation as going from $t_0$ to $t_1$ rather than vice versa. 
In complete analogy with our formalism above, we can therefore define a time-reversed Markov process
$\tilde{\M}$ whereby the future determines the past according to the time-reverse of \eq{MarkovEq}:
\beq{ReverseMarkovEq}
\p^{(0)}=\tilde{\M}\p^{(1)},
\eeq
where equations\eqn{MarkovMatrixEq2} and\eqn{qdefEq} get replaced  by
\beq{ReverseMarkovMatrixEq}
\tilde{M}_{ii'jj'}={p_{ii'jj'}\over p_{\cdot\cdot jj'}}
\eeq
and
\beq{qtildeDefEq}
\tilde{q}_{ii'jj'}=\tilde{M}^A_{ij}\tilde{M}^B_{i'j'}p^{(1)}_{jj'}\>.
\eeq
This time reversal symmetry doubles the number of $q$-options we could list in 
 \Tab{ComparisonTable} to six in total, augmenting  
 $q_{ii'jj'}$,  $q_{\cdot\cdot jj'}$ and  $q_{\cdot\cdot j\cdot}$ 
 by 
$\tilde{q}_{ii'jj'}$,  $\tilde{q}_{\cdot\cdot jj'}$ and  $\tilde{q}_{\cdot\cdot j\cdot}$.
In the interest of brevity, we have chosen to only list $\tilde{q}_{\cdot\cdot j\cdot}$, because of its
ability to kill $\phi$ for  efferent pathways --- the formulas for the two omitted options are trivially  analogous to those listed.

\subsection{Options for what to treat as known about the current state}
\label{ConditioningSec}

Above we listed options for which probabilities $p$ and $q$ to compare to compute $\phi$. 
To complete our specification of these probabilities, we need to choose between various options for our knowledge of the present state;
the three rightmost columns of \Tab{ComparisonTable} correspond to three interesting choices.

The first option is where the state is unknown, described simply by the probability distribution we have used above:
\beq{UnknownStateOptionEq}
p^{(0)}_{ij}=p_{ii'\cdot\cdot}
\eeq
This corresponds to us knowing $\M$, the mechanism by which the state evolves, but not knowing its current state $\x_0$. 
Note that a generic Markov process eventually converges to a unique stationary state $\p=\p^{(0)}=\p^{(1)}$ which, since it 
satisfies $\M\p=\p$, can be computed directly from $\M$ as the unique eigenvector whose eigenvalue is unity.\footnote{The only Markov processes that do not converge to a unique steady state are ones where $\M$ has more than one eigenvalue equal to unity; these form a set of measure zero on the set of all Markov processes. 
}
This means that if we consider a system that has been evolving for a significantly long time, its full two-time distribution $p_{ii'jj'}$ is determined by $\M$ alone; conversely, $p_{ii'jj'}$ determines $\M$ through \eq{MarkovMatrixEq2}.
Alternatively, if $p_{ii'jj'}$ is measured empirically from a time-series $\x_t$ which is then used to compute $\M$, 
we can use \eq{UnknownStateOptionEq} to describe our knowledge of the state at a random time. 

A second option is to assume that we know the initial probability distributions for  $\x^A_0$ and $\x^B_0$, but know nothing about any correlations between them.
This corresponds to replacing \eq{UnknownStateOptionEq} by the separable distribution
\beq{SeparableStateOptionEq}
p^{(0)}_{ij}=p_{i\cdot\cdot\cdot}p_{\cdot i'\cdot\cdot},
\eeq
and can be advantageous for $\phi$-measures that would conflate integration with initial correlations between the subsystems.

A third option, advocated by IIT \cite{tononi2008consciousness}, is to treat the current state as known:
\beq{KnownStateOptionEq}
p^{(0)}_{ij}=\delta_{ik}\delta_{i'k'},
\eeq
\ie, we know with certainty that the current state $\x_0=kk'$ for some constants $k$ and $k'$.
IIT argues that this is the correct option from the vantage point of a conscious system which, by definition, knows its own state.

A natural fourth option is a more extreme version of the first: treating the state not merely as unknown, with 
$p^{(0)}$ given by its ensemble distribution, but completely unknown, with a uniform distribution:
\beq{UniformStateOptionEq}
p^{(0)}_{ij}=\hbox{constant}.
\eeq
Although straightforward enough to use in our formulas, we have chosen not to include this option in \Tab{ComparisonTable} because it is rather inappropriate for most physical systems. For continuous variables such as voltages, it becomes undefined. For brains, such maximum-entropy states never occur: they would have typical neurons firing about half the time, corresponding to much more extreme ``on" behavior than during an epileptic seizure. The related option of consistently treating $\x_0^A$ as known but $\x_0^B$ as unknown when predicting $\x_1^A$ (and vice versa when predicting $\x_1^B$) corresponds to the noising factorization options described above. For further discussion of this, including so-called ``noising at the connection", see \cite{balduzzi2008integrated,seth2011causal,oizumi2014phenomenology}.

Finally, please note that if we choose to determine the past rather than the future 
(the ``p"-option from the previous section and \Tab{ComparisonTable}), then all the choices we have described should be applied 
to $p^{(1)}_{ij}$ rather than $p^{(0)}_{ij}$.

\begin{table*}
{\footnotesize
\renewcommand{\arraystretch}{1.8}
\begin{tabular}{|l|l|l|l|l|l|l|l|}
\hline
Code&Metric		&Definition							&Positivity	&Monotonicity	&Interpretability	&Tractability	&Symmetry\\
\hline
k&$d_{KL}(\p,\q)$	&$\sum\limits_\alpha p_\alpha \log {p_\alpha\over q_\alpha}$		&Y&Y&Y&Y&N\\
1&$d_1(\p,\q)$		&$\sum\limits_\alpha |p_\alpha-q_\alpha|$					&Y&Y&(Y)&N&Y\\
2&$d_2(\p,\q)$		&$\left[\sum\limits_\alpha(p_\alpha-q_\alpha)^2\right]^{1/2}$	&Y&Y&(N)&(Y)&Y\\
h&$d_H(\p,\q)$		&$\cos^{-1}\sum\limits_\alpha (p_\alpha q_\alpha)^{1/2}$		&Y&Y&Y&N&Y\\
s&$d_{SJ}(\p,\q)$	&$\left[\sum\limits_\alpha\left({p_\alpha\over 2}\log{2 p_\alpha\over p_\alpha+q_\alpha}+{q_\alpha\over 2}\log{2 q_\alpha\over p_\alpha+q_\alpha}\right)\right]^{1/2}$		&Y&Y&Y&N&Y\\
e&$d_{EM}(\p,\q)$	&$\min\limits_{f_{\alpha\beta}\ge 0}\sum\limits_{\alpha\beta}f_{\alpha\beta}d_{\alpha\beta}; \quad f_{\alpha\cdot}=p_\alpha, f_{\cdot\beta}=q_\beta$	&Y&Y&Y&N&Y\\
m&$d_{MD}(\p,\q)$	&See equations\eqn{dMDdefEq}-(\ref{IIstarDefEq})&Y&Y&Y&N&N\\
\hline
\end{tabular}
\caption{Different options for measuring the difference $d$ between two probability distributions $\p$ and $\q$:
Kullback-Leibler divergence $d_{KL}$, $L_1$-norm $d_1$,  $L_2$-norm $d_2$,  Hilbert-space distance $d_H$,  Shannon-Jensen distance $d_{SJ}$, 
Earth-Movers distance $d_{EM}$ and Mismatched Decoding distance $d_{MD}$.
These options correspond to the fourth superscript in $\phi$-measures such as $\phi^{ofuk}$.
In the text, we considered options where $\p$ and $\q$ had one, two or four indices, but in  this table, we have for simplicity combined all indices into a single Greek index $\alpha$.
\label{MetricTable}
}
}
\end{table*}

\subsection{Options for comparing probability distributions}

The options in the past three sections uniquely specify two probability distributions $\p$ and $\q$, and we want the integration $\phi$ to quantify how different they are from one another:
\beq{phiDefEq}
\phi\equiv d(\p,\q)
\eeq
for some distance measure $d$ that is larger the worse $\q$ approximates $\p$.
There are a number of properties that we may consider desirable for $d$ to quantify integration:
\begin{enumerate}
\item {\bf Positivity:} $d(\p,\q)\ge 0$, with equality if and only if $\p=\q$.
\item {\bf Monotonicity:} The more different $\q$ is from $\p$ in some intuitive sense, the larger $d(\p,\q)$ gets.
\item {\bf Interpretability:} $d(\p,\q)$ can be intuitively interpreted, for example in terms of information theory.
\item {\bf Tractability:} $d(\p,\q)$ is easy to compute numerically. Ideally, the optimal factorizations from \Sec{FactorizationSec} can be found analytically rather than through time-consuming numerical minimization. 
\item {\bf Symmetry:} $d(\p,\q)=d(\q,\p)$.
\end{enumerate}
Any distance measure $d$ meets the mathematical requirements of being a {\it metric} on the space of probability distributions if it obeys positivity, symmetry and the triangle inequality $d(\p,\q)\le d(\p,\rvec)+d(\rvec,\q)$.

\Tab{MetricTable} lists seven interesting probability distribution distance measures $d(\p,\q)$ from the literature together with their definitions and properties.
All these measures are seen to have the positivity and monotonicity, 
and all except the first are also symmetric and true metrics. 
We will now discuss them one by one in greater detail.

The distance $d_{KL}$ is the 
Kullback-Leibler divergence, and measures how many bits of information are lost when $\q$ is used to approximate $\p$, in the sense that if you developed an optimal data compression algorithm to compress data drawn from a probability distribution $\q$, it would on average require $d_{KL}(\p,\q)$ more bits to compress data drawn from a probability distribution $\p$ than if the algorithm had been optimized for $\p$ \cite{kullback1951information}. 
This has been argued to be the be the best measure because of its desirable properties related to information geometry \cite{amari2016information,griffith2014principled}.

$d_1$ and $d_2$ measure the distance between the vectors $\p$ and $\q$ using 
the $L_1$-norm and $L_2$-norm, respectively. The former is particularly natural for probability distributions $\p$, since they all have $L_1$ norm of unity: $d_1(0,\p)=p_.=1$. 
It is easy to see that 
$0\le d_1(\p,\q)\le 2$ and 
$0\le d_2(\p,\q)\le \sqrt{2}$.

The measure $d_H$ is the Hilbert-space distance: if, for each probability distribution, we define a corresponding wavefunction $\psi_i\equiv p_i^{1/2}$, then all wavefunctions lie on a unit hypersphere since they all have unit length: 
 $\langle\psi|\psi\rangle=p_.=1$.
The distance $d_H$ is simply the angle between two wavefunctions, \ie., the distance along the great circle on the hypersphere that connects the two, so $d_H(\p,\q)\le \pi/2$.
It is also the geodesic distance of the Fisher metric, hence a natural ``coordinate free" distance measure on the manifold of all probability distributions.

The measure $d_{SJ}$ is the Shannon-Jensen distance, whose square is defined as the average of the KL-divergences of the two distributions to their average: 
\beqa{SJeq}
d_{SJ}(\p,\q)^2&\equiv&{d_{KL}(\p,[\p+\q]/2)+d_{KL}(\q,[\p+\q]/2)\over 2}\nonumber\\
&=&S[(\p+\q)/2]-(S[\p]+S[\q])/2.
\eeqa
It is bounded by $0\le d_{SJ}\le 1$, satisfies the triangle inequality and is information-theoretically motivated \cite{endres2003new}.

The measure $d_{EM}$ is the Earth-Movers distance \cite{rubner1998metric}. If we imagine piles of earth scattered across the space $\x$, with $p(\x)$ specifying the fraction of the
earth that is in each location, then $d_{EM}$ is the average distance that you need to move earth to turn the distribution $p(\x)$ into $q(\x)$. The quantity $d_{ij}$ in the definition in \Tab{MetricTable}
specifies the distance between points $i$ and  $j$ in this space. For example, if $\x$ is a 3D Euclidean space, this may be chosen to be simply the Euclidean metric, while if $\x$ is a bit string, $d_{ij}$ may be chosen to be the $L_1$ ``Manhattan distance", \ie, the number of bit flips required to transform one bit string into another. 
IIT 3.0 argues that the earth mover's distance $d_{EM}$ is the most appropriate measure $d$ on conceptual grounds (whereas IIT 2.0 was still implemented using $d_{KL}$). Unfortunately, $d_{EM}$ rates poorly on the tractability criterion. It's definition involves a linear programming problem 
which needs to be solved numerically, and even with the fastest algorithms currently available, the computation grows faster than quadratically with the number of system states --- which in turn grows exponentially with the number of bits. For continuous variables $\x$, the number of states and hence the computational time is formally infinite. 



The measure $d_{MD}$ is based on ``mismatched decoding" as advocated by \cite{oizumi2016measuring}.
The distance measure $d_{MD}$ is defined not for all probability distributions, but for all  distributions over two variables, which we can write with two indices as $p_{ij}$:
\beq{dMDdefEq}
d_{MD}(\p,\q)\equiv I(\p)-\max_\beta I^*(\p,\q,\beta),
\eeq
where
\beqa{IIstarDefEq}
I(\p)&\equiv& -\sum_j p_{\cdot j}\log p_{\cdot j}
 +\sum_{ij} p_{ij}\log p_{j|i},\\
I^*(\p,\q,\beta)&\equiv& -\sum_j p_{\cdot j}\log\sum_i  q_{j|i}^\beta p_{i \cdot}
 +\sum_{ij} p_{ij}\log q_{j|i}^\beta,\nonumber
 \eeqa
 and the conditional distribution $q_{j|i}\equiv q_{ij}/q_{i\cdot}$.
Here $I(\p)$ is simply the mutual information between the two variables, since
combining \eq{IdefEq} with the conditional entropy definition from  \eq{ConditionalEntropyEq} gives the well-known equivalent expression for 
mutual information 
\beq{IdefEq2}
I(A,B)=S(A)-S(A|B).
\eeq 
$I^*(\p,\q,\beta)$ can be interpreted as the amount of information that one variable predicts about the other if the correct conditional distribution $p_{j|i}$ is 
replaced by a possibly incorrect one $q_{j|i}^\beta$ (renormalized to sum to unity) when making the prediction \cite{merhav1994information}.
This renormalization is strictly speaking unnecessary, because it cancels out between the two terms in $I^*(\p,\q,\beta)$.
Raising probabilities to positive powers $\beta$ has the effect of concentrating them (decreasing entropy) if $\beta>1$ and spreading them more evenly (increasing entropy) if $\beta<1$.
It can be shown that $I^*(\p,\q,\beta)\le I(\p)$ with equality for $\q=\p$ and $\beta=1$, and that $I^*(\p,\q,\beta)\ge 0$,  so one always has 
$0\le d_{MD}(\p,q)\le I(\p)$ \cite{merhav1994information}. Mismatched decoding can presumably be further generalized by replacing the maximization 
over powers $p^\beta$ by maximization over arbitrary monotonically increasing functions $f(p)$ that map the unit interval onto itself.

The integration measures of IIT3.0 have a more complex probability comparison that cannot be fully cast in the form of a simple function of $d(\p,\q)$: it makes the metric choice $d(\p,\q)=d_{\it EM}(\p,\q)$, but considers not only probability distributions for the whole system and a bipartition, but also for all possible subsets, providing an elaborate interpretation of the results in terms of ``conceptual structures" \cite{oizumi2014phenomenology}.


\section{Taxonomy results}
\label{TaxonomySec}

\subsection{Optimal factorization with $d_{KL}$}

Our taxonomy of integration measures is determined by four choices: of factorization, variable selection, conditioning and distance measure.
Although we have now explored these four choices one at a time, there are important interplays between them that we must examine.
First of all, the three optimal factorization options in \Tab{FactorizationTable} depend on what is being optimized, so let us now explore which of these optimizations are feasible and interesting to perform in practice and let us find out what the corresponding factorizations and $\phi$-measures are.

The mathematics problem we wish to solve is 
\beq{MinimizationEq}
\phi\equiv\min_{\M^A,\M^B} d(\p,\q)
\eeq
\ie, minimizing $d(\p,\q)$ over $\M^A$ and $\M^B$ given the constraints that 
$\M^A$ and $\M^B$ are markov Matrices: $M^A_{\cdot j}=1$, $M^B_{\cdot j'}=1$, $M^A_{ij}\ge 0$ and $M^B_{i'j'}\ge 0$.
\Tab{ComparisonTable} specifies the options for how $\p$ and $\q$ are computed and how $\q$ depends on $\M^A$ and $\M^B$, while \Tab{MetricTable} specifies the options for computing the distance measure $d$.
We enforce the column sum constraints using Lagrange multipliers, minimizing 
\beq{LagrangeEq}
{\cal L}\equiv d(\p,\q) -\sum_i \lambda_i (M^A_{\cdot i}-1)-\sum_{i'} \mu_{i'} (M^B_{\cdot i'}-1),
\eeq
and need to check afterwards that all elements of $\M^A$ and $\M^B$ come out to be non-negative (we will see that this is indeed the case).

As mentioned, numerical tractability is a key issue for integration measures. This means that it is valuable if  the Lagrange minimization can be rapidly solved analytically rather than slowly by numerical means, since this needs to be done separately for large numbers of possible system partitions. There is only  one $d$-option out of the above-mentioned five for which I have been able to solve the optimization over $\M$-factorizations analytically: the KL-divergence $d_{KL}$. The runner-up for tractability is $d_2$, for which everything can be easily solved analytically except for a final column normalization step, but the resulting formulas are cumbersome and unilluminating, falling foul of the interpretability criterion. Although $d_{KL}$ lacks the symmetry property, it has the above-mentioned positivity, monotonicity and interpretability properties, and we will now show that it also has the tractability property.

Let us begin with the $q$-options in the upper left corner of \Tab{ComparisonTable}, \ie, comparing the two-time distributions treating the present state as unknown.
Substituting \eq{qdefEq} into the definition of $d_{KL}$ from \Tab{MetricTable}
gives
\beqa{SeparableKLphiEq}
&&d_{KL}(\p,\q)=\sum_{ii'jj'} p_{ii'jj'} \log{p_{ii'jj'}\over p_{ii'\cdot\cdot} M^A_{ji}M^B_{j'i'}}=\\
&&S(\x_0)-S(\x)-\sum_{ij}p_{i\cdot j\cdot}\log M^A_{ji}-\sum_{i'j'}p_{\cdot i'\cdot j'}\log M^B_{j'i'},\nonumber
\eeqa
where the entropy for a random variable $\x$ with probability distribution $\p$ is given by Shannon's formula \cite{shannon1948mathematical}
\beq{SdefEq}
S(\x)=-\sum_i p_i\log p_i.
\eeq
To avoid a profusion of notation, we will often write as the argument of $S$ a random variable rather than its probability distribution.
For convenience, we will take all logarithms to be in base 2 for discrete distributions (so that entropies are measured in units of bits) and in base $e$ for continuous Gaussian distributions (so that equations get simpler).  In the latter case, where the entropy is based on the natural logarithm,  entropy is measured in ``nits" or ``nats" which equal $1/\ln 2\approx 1.44$ bits.

Substituting \eq{SeparableKLphiEq} into \eq{LagrangeEq} and requiring vanishing derivatives with respect to $M^A_{ij}$, $M^B_{i'j'}$, $\lambda_j$ and $\mu_{j'}$ shows that the solution to our minimization problem is 
\beq{OptimalSeparableMarkovEq}
M^A_{ji}={p_{i\cdot j\cdot}\over p_{i\cdot\cdot\cdot}},
\quad
M^B_{j'i'}={p_{\cdot i'\cdot j'}\over p_{\cdot i'\cdot\cdot}}.
\eeq
We recognize these equations as simply the Markov matrix estimator from \eq{MarkovMatrixEq} applied separately to subsystems A and B after marginalizing over the other system. 
Substituting this back into \eq{qdefEq} gives 
\beq{OptimalqEq}
q_{ii'jj'} = {p_{ii'\cdot\cdot} p_{i\cdot j\cdot} p_{\cdot i'\cdot j'}\over p_{i\cdot\cdot\cdot}p_{\cdot i'\cdot\cdot}}.
\eeq
Although the full probability distributions $q$ and $p$ typically differ, \eq{OptimalqEq} implies that three marginal distributions are identical:
$q_{i\cdot j\cdot}=p_{i\cdot j\cdot}$, $q_{\cdot i'\cdot j'}=p_{\cdot i'\cdot j'}$ and $q_{ij'\cdot\cdot}=p_{ij'\cdot\cdot}$.

Substituting \eq{OptimalqEq} back into
the definition of $d_{KL}$ 
gives the extremely simple result that the integration is 
\beqa{MarkovPhiEq}
\phi^{otuk}(\p) &=& 
\sum_{ii'jj'} p_{ii'jj'} 
\log{p_{ii'jj'}  p_{i\cdot\cdot\cdot}p_{\cdot i'\cdot\cdot} \over p_{ii'\cdot\cdot} p_{i\cdot j\cdot} p_{\cdot i'\cdot j'}}\nonumber\\
&=&I(\x^A,\x^B) - I(\x^A_0,\x^B_0),
\eeqa
where the mutual information between two random variables is given in terms of entropies by the standard definition
\beq{IdefEq}
I(\x^A,\x^B) \equiv S(\x^A)+S(\x^B)-S(\x).
\eeq

Since we will be deriving a large number of different $\phi$-measures that we do not wish to conflate with one another, we superscript each one with four code letters denoting the four taxonomical choices that define it. 
These letter codes are 
\begin{enumerate}
\item factorization: n/m/o/x/a
\item comparison: t/f/a/p
\item conditioning: u/s/k
\item measure: k/1/2/h/s/e/m
\end{enumerate}
and are defined in Tables~\ref{FactorizationTable}, \ref{ComparisonTable} and \ref{MetricTable}. For example, the integration measure $\phi^{otuk}$ from \eq{MarkovPhiEq} denotes optimized (o) factorization comparing the two-time (t) probability distributions with the current state unknown (u) and KL-divergence (k). Almost all measures discussed below will involve the k-measure (KL-divergence), so when this is the case we will typically drop this last index $k$ to avoid a confusing profusion of indices, for example writing $\phi^{otuk}=\phi^{otu}$.
For brevity, we will also define $\phi^M\equiv\phi^{otu}$, since we will be referring to this ``Markov measure" $\phi^{otu}$ many times below.

Although we derived this optimal factorization by comparing the two-time distribution (option t) for an unknown state (option u), an analogous calculation leads to the exact same optimal factorization for the options a$+$u, s$+$f and a$+$s. The option  t$+$s is undefined and the option
f$+$u gives messy equations I have been unable to solve analytically.
It is therefore reasonable to view \eq{OptimalSeparableMarkovEq}
as {\it the} optimal factorization when the state is unknown (option o), and for the remainder of this paper, we will simply define the o-option as using 
the factorization given by \eq{OptimalSeparableMarkovEq}.

Note that our result in \eq{MarkovPhiEq} involves a time-asymmetry, singling out $t_0$ rather than $t_1$ in the second term.
This is because we chose to interpret our Markov process as operating {\it forward} in time, determining the state at $t_1$ from the state at $t_0$.
As we discussed in \Sec{ComparisonSec}, we could equally well have done the opposite, using the Markov process $\tilde{M}$ operating {\it backward} in time, which would have yielded the alternative 
integration measure 
\beq{BackwardMarkovPhiEq}
\phi^{o\tilde{t}u}(\p) = I(\x^A,\x^B) - I(\x^A_1,\x^B_1).
\eeq
In practice, one usually estimates all statistical properties from a time-series that is assumed to be stationary.
This means that $I(\x^A_0,\x^B_0)=I(\x^A_1,\x^B_1)$, so that the these two integration measures become identical.

\subsection{Comparison with the  Ay/Barrett/Seth integration measures}

In the paper \cite{barrett2011practical} where Barrett \& Seth proposed their easier-to-compute integration measure $\phi^B$ (see below), they also 
mentioned an alternative measure that they termed ${\tilde\phi_E}$, defined by 
\beq{tildePhiEeq}
{\tilde\phi_E}\equiv S(\x_0^A|\x_1^A)+S(\x_0^B|\x_1^B)-S(\x_0|\x_1),
\eeq
where the {\it conditional entropy} of two variables $A$ and $B$ is defined by 
\beq{ConditionalEntropyEq}
S(A|B)\equiv S(A,B)-S(B).
\eeq
This measure had been introduced earlier by Ay \cite{ay2001information,ay2015information} in a context unrelated to IIT, under the name ``stochastic interaction", and was further discussed in \cite{oizumi2015unified,oizumi2016measuring}.
Applying equations\eqn{ConditionalEntropyEq} and\eqn{IdefEq} to \eq{tildePhiEeq} shows that
\beqa{BarrettEquivalenceEq}
{\tilde\phi_E}&=&S(\x^A)-S(\x_1^A)+S(\x^B)-S(\x_1^B)-S(\x)+S(\x_1)\nonumber\\
&=&I(\x^A,\x^B)-I(\x_1^A,\x_1^B)=\phi^{o\tilde{t}u},
\eeqa
\ie, that ${\tilde\phi_E}$ is identical to the time-reversed Markov measure $\phi^{o\tilde{t}u}$.
This equivalence provides another convenient interpretation of $\phi^{o\tilde{t}u}$:
as the average KL-divergence between (i) the probability distribution of the  past state $\x_0$ given the present state $\x_1$
and (ii) the product of these conditional distributions for the two subsystems.

It is also interesting to compare our result in \eq{MarkovPhiEq} with the popular integration measure 
\beq{BarrettPhiEq}
\phi^B(\p) = I(\x_0,\x_1) - I(\x^A_0,\x^A_1) - I(\x^B_0,\x^B_1)
\eeq
proposed by Barrett \& Seth \cite{barrett2011practical}.
The intuition behind this definition is to take the amount of information that a system predicts about its future and subtract of the information predicted by both of its subsystems.
Unfortunately, the result can sometimes go negative \cite{seth2011causal,oizumi2016measuring}, violating the desirable positivity property and making the $\phi_B$ difficult to interpret.
Consider the simple example of two independent bits that never change. 
If they start out perfectly correlated, then they will remain perfectly correlated, giving 
$I(\x_0,\x_1)=I(\x^A_0,\x^A_1)=I(\x^B_0,\x^B_1)=1$ and 
integrated information $\phi^B(\p)=-1$.

By substituting \eq{IdefEq} into equations\eqn{MarkovPhiEq} and\eqn{BarrettPhiEq}, we find that 
\beq{BarrettComparisonEq}
\phi^B(\p) =\phi^M(\p)-I(\x_1^A,\x_1^B).
\eeq
In other words, we can make the Barrett-Seth measure non-negative by adding back any final mutual information between the two subsystems.
When this is done, it becomes the integration measure we derived, therefore having a simple information-theoretic interpretation: 
it is the KL-divergence between the actual probability distribution $\p$ and the best separable approximation, which is guaranteed to be non-negative.

\subsection{Comparison with the mismatched decoding integration measure}

The measure $\phi^M$ is also closely related to the {\it mismatched decoding} measure $\phi^{MD}$ introduced in \cite{oizumi2016measuring}.
$\phi^{MD}$ makes the same taxonomical choices ``otu" as $\phi^M$ for the first three options:
optimal factorization (o), comparing full two-time distributions (t), and treating the past state as unknown (u).
However, it uses probability distance measure ``m" (mismatched decoding $d_{MD}$) instead of KL-divergence. We can therefore write this measure in our notation as
$\phi^{otum}=d_{MD}(p,q)$, where $q$ is the optimal factorization given by \eq{OptimalqEq}. Whether this factorization is also optimal 
in the sense of minimizing $d_{MD}(p,q)$ is not obvious.
 
The measure $\phi^M$ (or more specifically its time-reverse $\phi^{o\tilde{t}uk}$)
has been criticized in \cite{griffith2014principled,oizumi2016measuring} for being able to exceed the
mutual information $I(\x_0,\x_1)$ between the past and present: 
for example, if a two-bit system evolves from ``00'' to either ``00'' or ``11'' with equal probability, 
then $\phi^M = I(\x^A,\x^B)-I(\x^A_0,\x^B_0) = 1 - 0 = 1$ bit, even though $I(\x_0,\x_1)=0$.
This means that $\phi^M$ counts as a contribution to integration also correlated random noise added to both subsystems.
It is debatable whether this should count as integration: the ``con'' argument is that no information flows between the subsystems, while
the ``pro'' argument is that the two subsystems get linked by shared information flowing into both of them.

Both $\phi^M$ and $\phi^{MD}$ have intuitive bounds: 
$0\le \phi^M\le I(\x^A,\x^B)$ and $0\le \phi^{MD}\le I(\x_0,\x_1)$; these upper bounds correspond to the total 
mutual information across space and time, respectively.


\subsection{Optimal state-dependent factorization}

Let us now turn to factorization option ``x", optimized knowing the current state.
Consider some conscious observer (perhaps the system itself) who knows nothing about the system except its dynamics (encoded in $\M$) and its state at the present instant, encoded in $\x_0=kk'$. 
What can this observer say about the system state at earlier and later times? 
How integrated will this observer feel that the system is? 
To answer this question, we simply want to find the best approximate factorization of the conditional future state
$M_{jj'kk'}$ (or the past state $M_{kk'ii'}$), where $k$ and $k'$ are known constants. 

To gain intuition for this, let us temporarily write this conditional distribution as $p_{ii'}$, suppressing the known parameters $kk'$ for simplicity.
Given an arbitrary bivariate probability distribution $p_{ii'}$, what is best separarable approximation $q_{ii'}\equiv a_i b_{i'}$ in the sense that it minimizes
$d_{KL}(\p,\q)$? 
By minimizing $d_{KL}(\p,\q)$ using Lagrange multipliers, one easily obtains the long-known result that 
$a_i=p_{i.}$, $b_{i'}=p_{.i'}$ and $d_{KL}(\p,\q)=I$, the mutual information of $\p$.
In other words, even if we had never heard of marginal distributions or mutual information, we could derive them all from  $d_{KL}$:
the best factorization simply uses the marginal distributions, and the mutual information of a bivariate distribution is simply the KL-measure of how non-separable it is.

This means that the optimal factorization given $k$ and $k'$ is simply the one giving the marginal conditional distributions
\beq{OptimalFactorizationEq2}
M^A_{ji}={p_{kk' j\cdot}\over p_{kk'\cdot\cdot}},
\quad
M^B_{j'i'}={p_{kk'\cdot j'}\over p_{kk'\cdot\cdot}},
\eeq
and the corresponding integration is simply
\beq{ConditionedForwardMarkovPhiEq}
\phi^{xfkk} =I(\x_1^A,\x_1^B|\x_0).
\eeq
$\phi^{xtkk}$ is identical.
We can alternatively obtain this result directly from \eq{MarkovPhiEq} by noting that the $I(\x^A_0,\x^B_0)$-term vanishes now that the state $\x_0$ is known.

This result highlights a striking and arguably undesirable feature of measures based on the x-factorization option: they vanish for 
any deterministic system! If the system is deterministic and the present state $\x_0$ is known, then the future state $\x_1$ is also known, so all entropies in 
\eq{ConditionedForwardMarkovPhiEq} vanish and we obtain $\phi=0$.
With $\phi$-measures based on x-factorization,  the only source of integration is therefore correlated noise generated by the system.

\subsection{Minimizing integration on average}

Let us now turn to our final factorization option,  ``a", where we pick the state-independent factorization 
that minimizes integration on average.
Given the present state $\x_0=kk'$, let us compare the exact and approximate future probability distributions
\beqa{ConditionalComparisonEq}
p_{jj'}&=&P(\x_1=jj'|\x_0=kk')=\M_{kk'jj'},\\
q_{jj'}&=&P(\x_1^A=j|\x_0^A=k)P(\x_1^B=j'|\x_0^B=k')=\M^A_{kj}\M^B_{k'j'}\nonumber
\eeqa
by computing their KL-divergence $\phi=d_{KL}(p,q)$.
The answer clearly depends on the present state $kk'$, and we saw in the previous section what happens when we minimize separately for each state $kk'$.
Let us now instead average $d_{KL}(p,q)$ over all current states and find the state-independent factorization that minimizes this average:
\beqa{AverageDerivationEq}
\expec{d_{KL}(p,q)}
&=&\sum_{kk'}P(\x_0=kk')\>d_{KL}(p,q)|\x_0=kk'\nonumber\\
&=&\sum_{kk'}p_{kk'\cdot\cdot}\sum_{jj'}\M_{kk'jj'}\log{\M_{kk'jj'}\over \M^A_{kj}\M^B_{k'j'}}.
\eeqa
Substituting \eq{MarkovMatrixEq2} shows that this expression is identical to that from \eq{SeparableKLphiEq}, so minimizing it gives the exact same optimal factors
$\M^A$ and $\M^B$ and the exact same minimum $\phi$.
The comparison option ``t" gives the same result as well, 
so in conclusion, although they appear quite different from their definitions, the factorization options ``o" and ``a" are in fact identical.


\subsection{The full taxonomy} 

Now that we have derived the explicit form of all our factorization options, we can complete our integration measure classification.
Our taxonomy is determined by four choices: of factorization (n/m/o/x/a), variable selection (t/f/a/p), conditioning (u/s/k) and distance measure (k/1/2/h/s/e/m). Although this nominally gives $5\times 4\times 3\times 7=420$ different integration measures, most of these options turn out to be zero, undefined or identical to other options.\footnote{For noising factorizations (factorization options n and m), subsystem $B$ is randomized, so the only well-defined options are
$\phi^{nas*}$, $\phi^{nak*}$, $\phi^{nps*}$, $\phi^{npk*}$, 
$\phi^{mas*}$, $\phi^{mak*}$, $\phi^{mps*}$ and $\phi^{mpk*}$, 
where $*$ denotes any option for the distance measure.
For $o$-factorization, we find that 
$\phi^{oau*}=\phi^{oas*}=\phi^{opu*}=\phi^{ops*}=0$ and $\phi^{otk*}=\phi^{ofk*}$.
For $x$-factorization, $\phi^{xt**}$ is undefined and
one easily shows that $\phi^{xak*}=\phi^{xpk*}=0$, 
$\phi^{xau*}=\phi^{xas*}$
and
$\phi^{xpu*}=\phi^{xps*}$. We interpret k-conditioning as $\x_0$ being known for o-factorization and as $\x_0^A$ being known for noising factorizations, since the reverse options vanish and are undefined, respectively.
}

Whereas there are strong interactions between the factorization, variable selection and conditioning, we can freely choose any of the 7 distance measures independently of the other choices without changing whether $\phi$ vanishes or is well-defined.
We consider the option k (KL-divergence) by default below since it results in the simplest and most intuitive formulas; the formulas for the other options are straightforward to derive by combining Tables III, IV and~V. 
This leaves us with only the 21 separate options shown in \Tab{MeasureTable} to consider.
To provide intuition for these formulas, let us recapitulate key definitions in words:
\begin{itemize}
\item $\phi^M$ is the KL-divergence of the two-state probability distribution and the best separable approximation.

\item $\phi^{MD}$ is a measure of how much less information the present gives about the past if factorized dynamics is assumed.

\item $\phi^M_{kk'}$ is the KL divergence between (i) the future of the whole given the specific present state of the whole, and (ii) the product of this for the parts calculated separately. 

\item $\phi^{oak}$ is the KL divergence between (i) the distribution for the future state of subset A given the current state of A and (ii) the distribution for the future state of subset A given the current state of the whole system. 

\item $\phi^{opk}$ is $\phi^{oak}$ swapping ``future" for ``past". 

\item The subsequent ones are versions from above with different factorizations applied.

\end{itemize} 

\subsection{Which integration measures are best?}
 \label{BestSec}
 
 \Tab{PropertyTable} summarizes the desirable and undesirable traits for each of these integration measures, showing that merely a handful lack any major drawbacks. Let us now rate the various options in more detail.
 
For the choice of {\bf probability distance measure} (k/1/2/h/s/e/m), option ``e" (the Earth-Mover's distance $d_{EM}$ used in 
$\phi^{3.0}$ \cite{oizumi2014phenomenology}) remains an attractive candidate for discrete distributions with small number of bits, but is otherwise computationally unfeasible as we discussed above. All options in \Tab{PropertyTable} except $\phi^{3.0}$ and $\phi^{MD}$ therefore use option ``k" (the KL-divergence). Note that whether it is an advantage 
for the probability distance measure to be symmetric (as advocated in \cite{oizumi2014phenomenology}) depends on the 
interpretational context. For example, there is nothing asymmetric about the mutual information that ends up defining $\phi^{\rm M}$ in \Tab{MeasureTable}.

For the choice of {\bf factorization} (n/m/o/x/a), we can quickly dispense with option ``a'' (for being identical to ``o'') and option ``x'' (because it has the highly undesirable property of always vanishing for deterministic systems).
Which of the remaining options (n/m/o) is preferable depends on other choices.
If one wishes to use a distance measure other than the KL-divergence,  then the noising options ``n'' or ``m'' are computationally preferable, since the optimal factorization ``o'' can no longer be found analytically. Otherwise, ``m'' is arguably inferior to ``o'' because it is no simpler to evaluate and can overestimate the integration as described above.
If one has a philosophical preference for the factorization depending only on the mechanism $\M$ and not on any other information about state probabilities, then ``n'' is the only choice. If one wishes to consider continuous systems, on the other hand, ``n'' is undefined.
In summary, the best factorizations are therefore ``o'' and ``n'', depending ones preferences.
In practice, numerical experiments show that ``n'', ``m'' and ``o'' usually give quite similar $\phi$-values for a wide range of $\M$-matrices and probability distributions, so the choice between the three is a relatively minor one.

  
  
  

Turning now to the choice {\bf variable selection} and {\bf conditioning}, 
\Tab{PropertyTable} shows that many of the otherwise well-defined integration measures from 
\Tab{MeasureTable} have serious flaws.

\begin{figure}[phbt]
\centerline{\includegraphics[width=88mm]{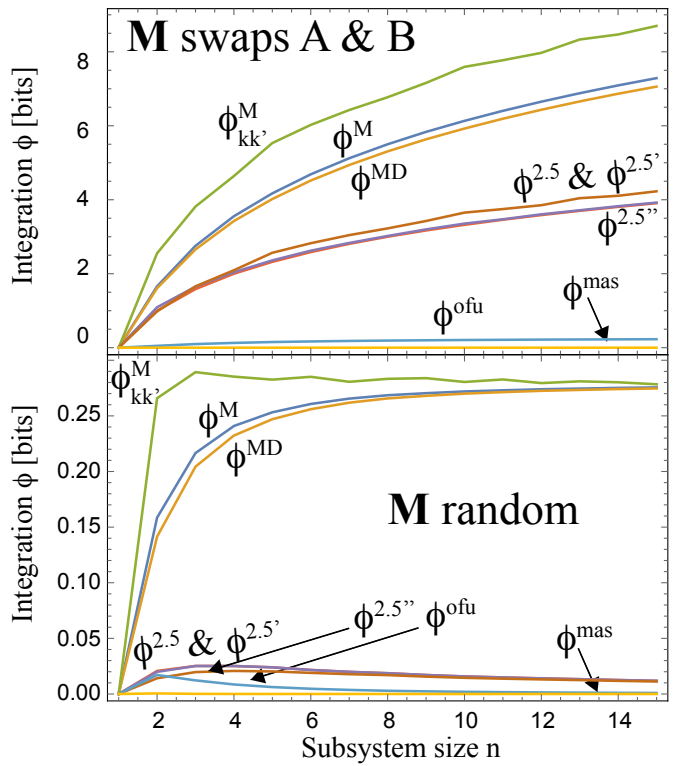}}
\caption{Numerical comparison of different integration measures, averaged over 3,000 random trials. In the bottom panel, all elements of $p$ are independently drawn from a uniform distribution and normalized to sum to unity.
In the top panel, only $p^{(0)}$ is randomly generated, and $\M$ is defined so as to swap the two subsystems, 
\protect\ie, 
$M_{jj'ii'}=\delta_{ij'}\delta_{i'j}$.
\label{IntegrationFig}
}
\end{figure}

Neither $\phi^{\rm ots}$ and $\phi^{\rm ofs}$ are guaranteed to vanish for separable systems, which means that we cannot in good conscience interpret them as measures of integration.
Numerical experiments show that $\phi^{\rm nas}$, $\phi^{\rm nps}$, $\phi^{\rm mas}$ and $\phi^{\rm nps}$ tend to be extremely small in practice ($\phi^{\rm mas}$ is plotted in \fig{IntegrationFig}). This is because they differ little from 
the corresponding measures using optimal factorization ($\phi^{\rm oas}$ and $\phi^{\rm ops}$), which always vanish. In other words, they are not really measures of  integration,  
merely measures of how suboptimal the factorizations $``n"$ and $``m"$ are.
For brevity, we have included merely three of these six flawed measures in \Tab{PropertyTable}.

\Fig{IntegrationFig} shows that $\phi^{\rm ofu}$ also tends to be much smaller than some other integration measures.
We can intuitively understand this by recalling that $\phi^{\rm oau}=0$, which means that optimal factorization lets us predict the future marginal distributions for A and B perfectly. Since $\phi^{\rm ofu}$ quantifies the inability of optimal factorization to predict the full future distribution, we expect that it will at most be of the order of  $I(\x_1^A,\x_1^B)$,  the extent to which this distribution is not separable (determined by its marginal distributions). 
For randomly generated probability distributions generated as in \Fig{IntegrationFig}), one can show that $I(\x_1^A,\x_1^B)\to 1-1/2\ln 2\approx 0.28$ bits in the limit where $n\to\infty$, and numerical experiments indicate that $\phi^{\rm ofu}$ is never much larger than this value for any $p$.

Dispensing with flawed/problematic $\phi$-measures narrows our list of remaining top candidates to merely nine:
$\phi^{\rm otu}$, $\phi^{\rm otum}$, $\phi^{\rm ofk}$, 
 $\phi^{\rm oak}$, $\phi^{\rm opk}$, 
 $\phi^{\rm nak}$, $\phi^{\rm npk}$, 
 $\phi^{\rm mak}$ and $\phi^{\rm mpk}$.
 Morover, the last six can be elegantly combined into merely three even better ones.
 As we discussed above, they have the advantage that they vanish for either afferent or efferent systems.
 
By following the prescription of \cite{tononi2008consciousness} and taking the minimum of two such complementary measures, we can construct an even better one that vanishes for {\it both} afferent and efferent systems.
All three of these improved measures are listed in \Tab{MeasureTable}.
The first is $\phi^{2.5}\equiv\min\{\phi^{\rm nak},\phi^{\rm npk}\}$.
We denote it ``2.5" because it combines attractive features of both IIT2.0 and IIT3.0: 
it starts with the $\phi^{\rm npk}$, which is precisely the IIT2.0 measure, and improves it by 
taking the minimum of cause/effect integration in the spirit of IIT3.0 (but retaining the KL-divergence of IIT2.0 instead of the harder-to-compute Earth-mover's distance of IIT3.0).
The second is $\phi^{2.5'}\equiv\min\{\phi^{\rm mak},\phi^{\rm mpk}\}$, which has the advantage of remaining defined even for continuous variables.
The third is $\phi^{2.5''}\equiv\min\{\phi^{\rm oak},\phi^{\rm opk}\}$, which uses the optimal factorization.

 \subsection{How large can $\phi$ get?}
 
 In summary, our taxonomy of $\phi$-measures produces merely a handful of truly attractive options: 
 $\phi^{2.5}$,  $\phi^{2.5'}$,  $\phi^{2.5''}$,  $\phi^{3.0}$, $\phi^{MD}$, $\phi^M$ and $\phi^M_{kk'}$.
\Fig{IntegrationFig} shows examples of what they evaluate to numerically.
The lower panel shows that for randomly generated probability distributions, none of them exceed 
$1-1/2\ln 2\approx 0.28$ bits on average, which as mentioned above is the mutual information in a random bivariate distribution.
However, $\phi^{2.5}$,  $\phi^{2.5'}$,  $\phi^{2.5''}$,  $\phi^M$, $\phi^{MD}$ and $\phi^M_{kk'}$ can get arbitrarily large for some systems, as illustrated in the top panel, growing logarithmically with the size $n$ of the subsystems A and B.
In other words, the maximum integration is of the order of the number of subsystem bits. 
For the example shown where the dynamics merely swaps the two subsystems, we obtain $\phi^{2.5}=\log_2 n$, because noising gives $M^A = 1/n$, $q = 1/n^2$ and $p$ is a Kronecker $\delta$. 
$\phi^{M}$, $\phi^{MD}$ and $\phi^M_{kk'}$ are seen to give about twice the integration for this example.

Note that although this dynamics $\M$ that merely swaps the subsystems has such a large $\phi$-value only for this particular cut that separates the systems being swapped. 
Consider, for example, a system of four bits labeled 1, 2, 3 and 4, where the dynamics swaps 1 with 3 and 2 with 4.
There is a different cut where $\phi=0$: simply define the new subsystems A' and B' to be the first and second halves of the A and B-systems, \ie, 
$A'={1,3}$ and $B'={2,4}$. The swapping is now carried out internally within A' and B', revealing that there is no integration and upper-case $\Phi=0$.

However, there are plenty of systems for which even the true integration $\Phi$ grows like the number of subsystem bits, $\log_2 n$.
A simple example accomplishing this (in the spirit of the random coding example in \cite{tegmark2014consciousness}) is when
the $n^4$ probabilities $p_{ii'jj'}$ are all set to zero except for a randomly selected subset of $n^2$ of them that are set to $1/n^2$.
Now $\phi^M\sim\log_2 n$ even when minimized over all bipartitions of the $2\log_2 n$ bits in the system.\footnote{For this example, 
we have $S(\x)=\log_2 n^2=2\log_2 n$. The marginal distributions for $\x^A$, $\x^B$, $\x_0^A$ and $\x_0^B$ are all rather uniform, 
with entropy on average less than a bit from the value for a uniform distribution, giving 
$S(\x^A)\sim S(\x^B)\sim\log_2 n^2$, 
$S(\x_0^A)\sim S(\x_0^B)\sim\log_2 n$, 
$I(\x^A,\x^B)=S(\x^A)+S(\x^B)-S(\x)\sim 2\log_2n$,
$I(\x_0^A,\x_0^B)=S(\x_0^A)+S(\x_0^B)-S(\x_0)\sim 0$ and therefore 
$\phi^M=I(\x^A,\x^B)-I(\x_0^A,\x_0^B)\sim 2\log_2 n\sim\log_2 n$.
}

\begin{figure}[phbt]
\centerline{\includegraphics[width=88mm]{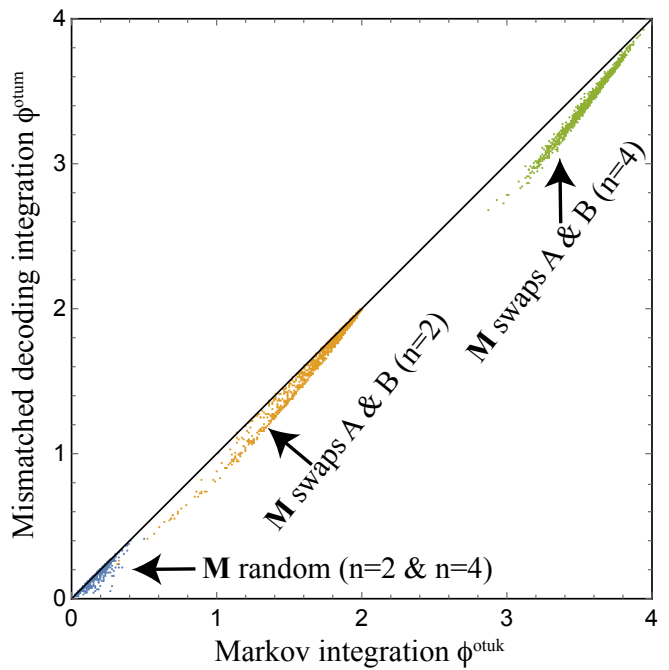}}
\caption{Numerical comparison of the two measures $\phi^{otuk}$ and $\phi^{ofum}$ for 3,000 random trials, generated the same way as in \fig{IntegrationFig}.
The two measures are seen to be rather similar for these examples, and to satisfy the inequality $\phi^{ofum}\le\phi^{otuk}$.
\label{MismatchedDecodingFig}
}
\end{figure}

\Fig{MismatchedDecodingFig} shows that the measures $\phi^M$ and $\phi^{MD}$ can sometimes be quite similar:
they give numerically similar values for the 3,000 random examples shown. Moreover, they appear to satisfy the inequality
$\phi^{ofum}\le\phi^{otuk}$. Further examination shows that for these these random examples, the $\beta$-complication in 
\eq{dMDdefEq} makes essentially no perceptible difference in practice, in the sense that the computation of
$\phi^{MD}$ can be accurately accelerated by setting $\beta=1$ rather than minimizing over it.
However, \cite{oizumi2016measuring} shows that there are real-world cases where $\beta$ is far from unity and also where
$\phi^{MD}\ll\phi^{M}$, particularly when noise correlations dominate over causal correlations.
To understand this, consider the extreme case of two perfectly correlated bits that are independently randomized by both time 0 and time 1, so that
$\x_0^A=\x_0^B$ and $\x_1^A=\x_1^B$, with no correlation between the two times.
Then $\phi^{MD}=0$ whereas $\phi^M=I(\x^A,\x^B) - I(\x^A_0,\x^B_0)=2-1=1$, which is arguably undesirable.

 
 %


\section{The $n\to\infty$ limit of continuous variables}
\label{GaussianSec}

All our previous results are fully general, applying regardless of whether the variables are discrete 
(such as bits that equal zero or one) or continuous (such as voltages or other variables measured in 
fMRI, EEG, MEG or electrophysiology studies). We can view the latter as the $n\to\infty$ limit of the former, since a single real number can be represented as an infinite string of bits.
In this section, we will focus on the continuous case and see how our previous formulas can be greatly simplified by assuming Gaussianity.
We therefore replace $i$, $i'$, $j$ and $j'$ in all our formulas by 
$\x_0^A$, $\x_0^B$, $\x_1^A$ and $\x_1^A$, respectively, and replace all sums by integrals.

\subsection{How Gaussianity gives linearity}

To make things tractable, we will make one strong but very useful assumption: that $\x$ has a Gaussian distribution.
The most general $d$-dimensional multivariate Gaussian distribution is parametrized by its mean vector $\m\equiv\expec{\x}$ and covariance matrix $\T\equiv\expec{\x\x^t}-\m\m^t$ and takes the form
\beq{GaussianpEq}
g[\x;\m,\T]\equiv{1\over (2\pi)^{d/2} |\T|^{1/2}}e^{-{1\over 2}(\x-\m)^t\T^{-1}(\x-\m)},
\eeq
so we are making the assumption that there is some $\m$ and $\T$ such that $p(\x)=g(\x;\m,\T)$.
Let us write $\m$ and $\T$ as 
\beq{Meq2}
\m=
\left(
\begin{tabular}{c}
$\m_0$\\
$\m_1$
\end{tabular}
\right),
\quad
\T=
\left(
\begin{tabular}{cc}
$\C_0$&$\B$\\
$\B^t$&$\C_1$
\end{tabular}
\right),
\eeq
where $\m_i$ and $\C_i$ are the mean and covariance of $\x_i$, respectively.

Interpreting the sum in the denominator of \eq{MarkovMatrixEq2} as an integral and evaluating it\footnote{
 The following well-known matrix identities are useful in the derivation of  this and other matrix results in this paper:
 \beq{BlockDeterminantEq}
 \left|
\begin{tabular}{cc}
\A&\B\\
\C&\D
\end{tabular}
\right|
= |\A\D-\A\C\A^{-1}\B|,
\eeq
 \beq{BlockInverseEq}
\left(\!\!
\begin{tabular}{cc}
\A&\B\\
\C&\D
\end{tabular}
\!\!\right)^{-1}\!\!\!\!\!\!
=
\left(\!\!
\begin{tabular}{cc}
$[\A-\B\D^{-1}\C]^{-1}$&$-\A^{-1}\B[\D-\C\A^{-1}\B]^{-1}$\\
$[\D-\C\A^{-1}\B]^{-1}\C\A^{-1}$&$[\D-\C\A^{-1}\B]^{-1}$
\end{tabular}
\!\!\right),
\eeq
\beq{WoodburyEq}
[\A+\B\D^{-1}\C]^{-1}=\A^{-1}-\A^{-1}\B[\D+\C\A^{-1}\B]\C\A^{-1}.
\eeq
}
gives
\beq{ContinuousMarkovEq}
\M(\x_1,\x_0)=g[\x_1;\m_1+\A(\x_0-\m_0),\SS],
\eeq
where
\beqa{Aeq}
\A&\equiv&\B^t\C_0^{-1},\\
\SS&\equiv&\C_1-\B^t\C_0^{-1}\B=\C_1-\A\C_0\A^t.\label{Seq}
\eeqa
This encodes the well-known result that the conditional distribution $\x_1|\x_0$ for Gaussian variables is Gaussian with 
mean $\m_1+\B\C_0^{-1}(\x_0-\m_0)$ and covariance matrix $\C_1-\B^t\C_0^{-1}\B$.
These equations embody a remarkable simplicity that we can exploit. 
First of all, the covariance matrix $\SS$ is independent of $\x_0$, which allows us to interpret 
$\x_1$ as simply a function of $\x_0$ plus a random noise vector $\n$ that is independent of $\x_0$. 
Second, this function is affine, involving simply a linear term plus a constant.
In other words, we can write 
\beq{LinearEq}
\x_1=\m_1+\A(\x_0-\m_0)+\n,
\eeq
where the noise vector $\n$ satisfies
\beq{nEq}
\expec{\n}=0,\quad\expec{\n\x^t}=0,\quad\expec{\n\n^t}=\SS.
\eeq
It is worth reflecting on how remarkable this is, since it is easy to overlook. 
The future state $\x_1$ of a system can depend on the present state $\x_0$ in some arbitrarily complicated non-linear way. Moreover, for a generic Markov process, the scatter of $\x_1$ around its mean will depend strongly on $\x_0$. Yet as long as all probability distributions are Gaussian, which is often a useful approximation for laboratory data, both of these complications vanish and we are left with the simple linear dynamics of \eq{LinearEq}.


\subsection{Autoregressive processes}

Let us now briefly review the formalism of so-called autoregressive processes and how it relates to our problem at hand.
A simple special case of the above is where the random process is {\it stationary}, \ie, where the statistical properties are independent of time. This implies that $\m_i=\m$ and $\C_i=\C$ for some $\m$ and $\C$ that are independent of $i$.
For a stationary process, it is convenient to redefine new zero-mean variables $\x_i'\equiv\x_i-\m$. Dropping the prime for simplicity, this allows us to rewrite \eq{LinearEq} as
\beq{LinearEq2}
\x_{i+1}=\A\x_{i}+\n_i,
\eeq
where the noise vectors $\n_i$ have vanishing mean and vanishing correlations between different times, \ie, $\expec{\n_i\n_j^t}=\delta_{ij}\SS$.
The covariance matrix between vectors at two subsequent times is therefore
\beq{Teq1}
\expec{\x\x^t}\equiv \left(
\begin{tabular}{c@{\hskip 5mm}c}
$\C$&$\C\A^t$\\
$\A\C$&$\A\C\A^t+\SS$
\end{tabular}
\right),
\quad
\x\equiv
\left(
\begin{tabular}{c}
$\x_0$\\
$\x_1$
\end{tabular}
\right).
\eeq
Even if the random process is not stationary initially, it will eventually converge to a stationary state where covariance is time-independent as long as all eigenvalues of $\A$ have magnitude below unity, so that memory of the past gets exponentially damped over time.
Once the covariance has become time-independent, \eq{Teq1} implies that $\C=\A\C\A^t+\SS$. 
This is known as the Lyapunov equation, and is readily solved by special-purpose techniques or, rapidly enough, by simply iterating it to convergence.
If we write the covariance matrix $\expec{\x\x^t}$ measured from actual time series data as
\beq{Teq2}
\T\equiv\expec{\x\x^t}=
\left(
\begin{tabular}{cc}
$\C$&$\B$\\
$\B^t$&$\C$
\end{tabular}
\right),
\eeq
then equating it with \eq{Teq1} lets us compute the matrices we need from the data:
\beqa{Aeq2}
\A&=&\B^t\C^{-1},\\
\SS&=&\C-\A\C\A^t=\C-\B^t\C^{-1}\B.\label{Seq2}
\eeqa
These equations hold regardless of whether the probability distributions are Gaussian or not. 
If the noise $\n$ is Gaussian, then all distributions will be Gaussian in the steady state, so this is an alternative way of deriving 
equations~\eqn{ASeq} and~\eqn{Seq} (without the subscripts).

In \Sec{ComparisonSec}, we saw how we can equally well interpret our system as a Markov process operating backward in time, where the future causes the past. Repeating the above derivation for this case, we can write 
\beq{BackwardLinearEq}
\x_{i-1}=\At\x_{i}+\n_i,
\eeq
where
\beqa{AtildeEq}
\At&=&\B\C^{-1}=\SS\A^t\SSt^{-1},\\
\SSt&=&\C-\B\C^{-1}\B^t=\C-\At\C\At^t,\label{StildeEq}\\
&=& [\C^{-1}+\A^t\SS^{-1}\A]^{-1}=\C-\C\A^t\C^{-1}\A\C.\nonumber
\eeqa
Although the matrices $\SS$ and $\SSt$ are different, it is easy to prove that their determinants are identical, which means that the conditional entropy is the same both forward and backward in time.

\subsection{Optimal factorization}

In summary, a Markov process $\p_1=\M\p$ can be described much more simply when all probability distributions are Gaussian: instead of keeping track of the infinite-dimensional Markov matrix $\M$ or the infinite-dimensional rank-4 tensor $\p$ 
(both of which have as indices the four continuous variables $\x_0^A$, $\x_0^B$, $\x_1^A$, $\x_1^B$), 
we merely need to keep track of the $2n\times 2n$ covariance matrix $\T$, from which we can compute and quantify the deterministic and stochastic parts of the dynamics as the matrices $\A$ and $\SS$, respectively.

Let us now translate the rest of our results from our integration taxonomy into this simpler formalism.
To separate out the effects occurring within and between the subsystems A and B, let us name the corresponding blocks of the $\A$-matrix and the
matrix $\T\equiv\expec{\x\x^t}$ from \eq{Meq2} as follows:
\beqa{ApartsDefEq}
\A&=&
\left(
\begin{tabular}{ll}
$\A_A$		&$\A_{AB}$\\
$\A_{BA}$	&$\A_B$
\end{tabular}
\right)\!\!,\\
\T&=&
\left(
\begin{tabular}{llll}
$\C_A$		&$\C_{AB}$		&$\B_A$		&$\B_{AB}$\\
$\C_{AB}^t$	&$\C_B$			&$\B_{BA}$	&$\B_B$\\
$\B_A^t$		&$\B_{BA}^t$		&$\C_A$		&$\C_{AB}$\\
$\B_{AB}^t$	&$\B_B^t$		&$\C_{AB}^t$	&$\C_B$
\end{tabular}
\right)\!\!,
\>\>
\x=
\left(
\begin{tabular}{l}
$\x^A_0$\\
$\x^B_0$\\
$\x^A_1$\\
$\x^B_1$
\end{tabular}
\right)\!\!.
\eeqa
Analogously to how \eq{MarkovMatrixEq2} gave us \eq{ContinuousMarkovEq}, 
\eq{OptimalSeparableMarkovEq} now gives the optimal factorization
\beqa{ContinuousMarkovEq2}
\M^A(\x_1^A,\x_0^A)&=&g[\x_1^A;\Ah_A\x_0^A,\SSh_A],\\
\M^B(\x_1^B,\x_0^B)&=&g[\x_1^B;\Ah_B\x_0^B,\SSh_B],
\eeqa
where
\beqa{ASeq}
\Ah_A&\equiv&\B_A^t\C_A^{-1},\quad \SSh_A\equiv\C_A-\B_A^t\C_A^{-1}\B_A,\\
\Ah_B&\equiv&\B_B^t\C_B^{-1},\quad \SSh_B\equiv\C_B-\B_B^t\C_B^{-1}\B_B.
\eeqa
In other words, the ``o''-factorization approximates $\x_1=\A\x_0+\n$ by
\beq{xhDefEq}
\xh_1\equiv
\left(
\begin{tabular}{l}
$\xh^A_1$\\
$\xh^B_1$\\
\end{tabular}
\right)
\equiv\Ah\x_0+\nh,
\quad
\Ah\equiv
\left(
\begin{tabular}{cc}
$\Ah_A$	&$0$\\
$0$		&$\Ah_B$
\end{tabular}
\right),
\eeq
where the noise vector $\nh$ has zero mean and covariance matrix
\beq{ShDefEq}
\SSh\equiv
\left(
\begin{tabular}{cc}
$\SSh_A$	&$0$\\
$0$		&$\SSh_B$
\end{tabular}
\right).
\eeq
We see that tensor factorization in the previous section now corresponds to the matrices
$\A$ and $\SS$ being block-diagonal.

\subsection{Noising factorization}

\Eq{LinearEq2} tells us that
\beq{NoisingxEq}
\left(
\begin{tabular}{c}
$\x^A_1$\\
$\x^B_1$
\end{tabular}
\right)=\A\x_0+\n=
\left(
\begin{tabular}{c}
$\A_A\x^A_0+\A_{AB}\x^B_0+\n^A$\\
$\A_B\x^B_0+\A_{BA}\x^A_0+\n^B$
\end{tabular}
\right).
\eeq
The idea with noising is to take the terms $\A_{AB}\x^B_0$ and $\A_{BA}\x^A_0$ and reinterpret them not as signal but as noise, with zero mean and uncorrelated with anything else.
The noising option ``n'' is unfortunately undefined for this continuous-variable case, because it says to use a uniform distribution for these noised versions of $\x_0^A$ and $\x_0^B$, which has infinite variance and hence gives, \eg,  $\expec{\x^B_0{\x^B_0}^t}=\infty$ when $\x_0^B$ is noised.
The mild noising option ``m'', however, remains well-defined, saying to use the actual distributions for these noised versions of $\x_0^A$ and $\x_0^B$, hence giving $\expec{\x^A_0{\x^A_0}^t}=\C_A$ and $\expec{\x^B_0{\x^B_0}^t}=\C_B$ when these variables are noised.

Computing the first and second moments of \eq{NoisingxEq} therefore tells us that 
``m''-factorization approximates $\x_1=\A\x_0+\n$ by
\beq{NoisingxhDefEq}
\xb_1\equiv
\left(
\begin{tabular}{l}
$\xb^A_1$\\
$\xb^B_1$\\
\end{tabular}
\right)
\equiv\Ab\x_0+\nb,
\quad
\Ab\equiv
\left(
\begin{tabular}{cc}
$\A_A$	&$0$\\
$0$		&$\A_B$
\end{tabular}
\right),
\eeq
where the noise vector $\nb$ has zero mean and covariance matrix
\beq{NoisingShDefEq}
\SSb\equiv
\left(
\begin{tabular}{cc}
$\SS_A+\A_{AB}\C_B\A_{AB}^t$	&$0$\\
$0$							&$\SS_B+\A_{BA}\C_A\A_{BA}^t$
\end{tabular}
\right).
\eeq
Note that in contrast to the ``o''-factorization of \eq{xhDefEq}, 
the ``m''-factorization has no tildes on the $\A_A$ and $\A_B$-matrices in \eq{NoisingxhDefEq}.

\subsection{Results}

We now have all the tools we need to derive the Gaussian versions of the $\phi$-formulas in \Tab{MeasureTable}.
Starting with \eq{IdefEq}, interpreting the sum in \eq{SdefEq} as an integral and performing it when $p$ is the Gaussian distribution of  \eq{GaussianpEq}
gives the well-known formula
\beq{GaussianIeq}
I(\x_A,\x_B)={1\over 2}\log{ |\T_A|\,|\T_B|\over |\T|}
\eeq
for the mutual information between two multivariate Gaussian random variables.
This immediately gives the five matrix formulas for $\phi^{\rm M}$, $\phi^{\rm B}$, $\phi^{\rm ots}$, $\phi^{\rm ofs}$ and $\phi^{\rm xfk}$ in the right column of  \Tab{MeasureTable}.
The second version listed for $\phi^{\rm B}$ is also given in  \cite{barrett2011practical}.

Starting with the  KL-divergence definition $d_{\rm KL}(p,q)\equiv\sum_i p_i\log{p_i\over q_i}$ from \Tab{MetricTable}, we again interpret the sum as an integral and use 
\eq{GaussianpEq}. This gives the well-known formula 
\beqa{GaussianKLeq}
&&D_{KL}(f_p,f_q)=\nonumber\\
&&{1\over 2}\left[\Delta\m^t\C_q^{-1}\Delta\m + \tr \C_q^{-1}\C_p +\ln{|\C_q|\over|\C_p|}-n\right]
\eeqa
for the KL-divergence between two Gaussian probability distributions $f_p$ and $f_q$ with means $\m_i$ and covariance matrices $\C_i$ $(i=p,q)$,  
where $\Delta\m\equiv\m_p-\m_q$.
The first term in \eq{GaussianKLeq} thus represents the mismatch between the means and the
remainder (which is also guaranteed to be nonnegative) represents the mismatch between the covariances.

For $\phi^{\rm ofu}$, the future distribution $p(\x_1)$ with mean zero and covarance matrix $\C$ is approximated by
the distribution $q(\x_1)$ that has mean zero and covariance matrix $\Ah\C\Ah^t+\SSh$, which follows from 
equations\eqn{xhDefEq} and\eqn{ShDefEq}.
Substituting these means and covariance matrices into \eq{GaussianKLeq} gives the matrix formula for $\phi^{\rm ofu}$ in \Tab{MeasureTable}.
For $\phi^{\rm mas}$,  both means again vanish, but now the future distribution $p(\x_1^A)$ has covariance matrix $\C_A$ while the approximation
$q(\x_1^A)$  has covariance matrix $\SS_A+\A_A\C_A\A_A^t+\A_{AB}\C_B\A_{AB}^t $, which follows from 
equations\eqn{NoisingxhDefEq} and\eqn{NoisingShDefEq}.

For the remaining options in \Tab{MeasureTable}, \ie,
$\phi^{\rm ofk}$, $\phi^{\rm oak}$, $\phi^{\rm opk}$, $\phi^{\rm mak}$ and $\phi^{\rm mpk}$, the means do not vanish, since they 
reflect information about the known state.
For $\phi^{\rm ofk}$, the future distribution $p(\x_1)$ with mean $\A\x_0$ and covariance matrix $\SS$ is approximated by
the distribution $q(\x_1)$ that has mean $\Ah\x_0$ and covariance matrix $\SSh$, so \eq{GaussianKLeq} gives the matrix formula for $\phi^{\rm ofk}$ in the table.
For $\phi^{\rm oak}$,  the future distribution $p(\x_1^A)$ has mean $\A_A\x_0^A+\A_B\x_0^B$ and covariance matrix $\SS$, while the approximation
$q(\x_1^A)$  has mean $\Ah_A\x_0^A$ and  covariance matrix $\SSh_A$.
Finally, for $\phi^{\rm mak}$,  the future distribution 
$p(\x_1^A)$ with mean $\Ah_A\x_0^A$ and  covariance matrix $\SSh_A$ is approximated by 
$q(\x_1^A)$ with mean $\Ab_A\x_0^A$ and  covariance matrix $\SSb_A$. 
The time-reversed measures 
$\phi^{\rm opk}$, $\phi^{\rm mps}$ and $\phi^{\rm mpk}$ are identical to 
$\phi^{\rm oak}$, $\phi^{\rm mas}$ and $\phi^{\rm mak}$, but with $\A$ and $\SS$ replaced by their time-reversed versions $\At$ and $\SSt$ from \eq{AtildeEq}.

Substituting the above Gaussian formulas into equations~\eqn{dMDdefEq} and~\eqn{IIstarDefEq} gives
\beq{GaussianPhiMDeq}
\phi^{MD}={\ln{|\Ch|\over|\SSh|}
+\tr\left[\Ch^{-1}\C-\SSh^{-1}\SS-\Ap^t\SSh^{-1}\Ap\C\right]\over 2},
\eeq
where $\Ap\equiv\A-\Ah$ and $\Ch\equiv\Ah\C\Ah^t+\SSh$ for the simple but important case $\beta=1$.

\section{Graph-theory approximation to make computations feasible}
\label{ApproximationSec}

\subsection{The problem}

The $\phi$-formulas for discrete variables in the left column of \Tab{MeasureTable} require working with the
$n\times n$ matrix $\M$, where $n=2^b$ for a system of $b$ bits.
In other words, the time to evaluate $\phi$ for a given cut grows exponentially with the system size $b$, which becomes computationally prohibitive even for modest system sizes such as 100 bits --- let alone the set of neurons in the human brain with $b\sim 10^{11}$. Even 300 bits give $n$ greater than the number of particles in our universe.

When the system state is described not by bits but continuous variables  (such as voltages or other variables measured in 
fMRI, EEG, MEG or electrophysiology studies), things get even worse, since represending even a single variable requires an infinite number of bits. However,  \cite{barrett2011practical} pointed out that the Gaussian approximation radically simplifies things, and we saw in \Sec{GaussianSec} how $\phi$ can then be computed dramatically faster.
Not only does the infinity problem go away for most measures in \Tab{MeasureTable}, but the formulas in the right column are exponentially faster to evaluate than those in the left column even when each bit is replaced by a separate real number! This is because if there are $b$ real numbers, the $n\times n$ matrix $\T$ has $n=2b$, not $n=2^b$. 
This means that $\phi$ can now be computed in polynomial time, more specifically $O(b^3)$ time, since 
the slowest matrix operations in \Tab{MeasureTable} scale as $O(n^3)$.

Unfortunately, even after this exponential speedup in computing $\phi$, computing the upper-case version $\Phi$ is still exponentially slow. This is because $\Phi$ is the minimum of $\phi$ over the exponentially many ways of splitting the system into two parts. Even if we limit ourselves to symmetric bipartitions, the number of ways to split an even number  $n$ elements into two parts of size $n/2$ is 
\beq{StirlingEq}
\left({n\atop n/2}\right)={n!\over (n/2)!^2}
\approx \sqrt{2\over\pi n} 2^n,
\eeq
where we have used Stirling's approximation $n!\approx\sqrt{2\pi n}(n/e)^n$.
In other words, examining all symmetric bipartitions is pretty much as exponentially painful as examining all $2^n$ bipartitions, because most bipartitions are close to symmetric.

\begin{figure}[phbt]
\centerline{\includegraphics[width=88mm]{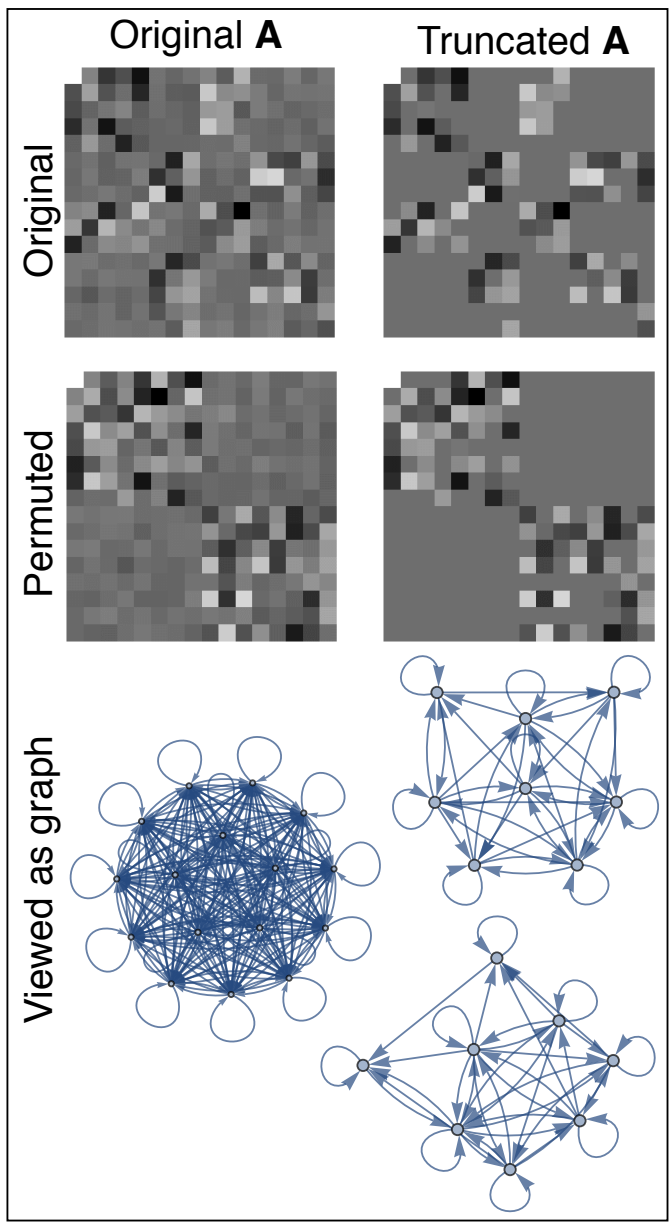}}
\caption{Illustration of our fast $\Phi$-approximation for an $n=16$ example. 
The structure of the $\A$-matrix can be visualized either as a grid (top four examples) where each pixel color shows the value of the corresponding element $A_{ij}$ ranging from the smallest (black) to the largest (white), or as a graph (bottom examples) showing all non-zero matrix elements.
Both of the matrices on the left correspond to the same graph below them, and both of the matrices on the right correspond to the same (disconnected) graph below them.
Our method zeros all matrix elements $|A_{ij}|<\epsilon$ below the threshold $\epsilon$ that makes the largest connected graph component involve merely half of the elements, which in the matrix picture means that there is a permutation of the elements (rows and columns) rendering the matrix block-diagonal (middle right). Whereas it would take exponentially long to try all matrix permutations, graph connectivity can be determined in polynomial time, thus enabling us to rapidly find a good approximation for the ``cruelest cut'' bipartition.
\label{GraphsFig}
}
\end{figure}

\subsection{An approximate solution}

Being able to compute $\Phi$ approximately is clearly better than not being able to compute it at all. 
In this spirit, let us explore an approximation that exponentially accelerates the computation of $\Phi$.
Starting with the linear dynamics $\x_{i+1}=\A\x+\n$
from \eq{LinearEq2},
let us motivate our approximation by considering the case where the noise is $\n$ uncorrelated (where $\SS$ is diagonal) so that it introduces no correlations between the two systems, regardless of the cut. This means that the only source of integration can be the $\A$-matrix transferring information between the two subsystems.
Let us visualize this information flow as a directed graph (\Fig{GraphsFig}, bottom), where each node represents a variable 
$i$ and each edge represents a non-zero element $A_{ij}$, \ie, non-zero information flow from element $j$ to element $i$. 
If this graph consists of two disconnected parts A and B of equal size, as in the lower right corner of \Fig{GraphsFig}, then we clearly have $\Phi=0$, since there is no information flow and hence no integration between these two parts. In other words, if we permute the elements so that all elements of A precede all elements of B, the matrix $\A$ becomes block-diagonal (\Fig{GraphsFig}, middle right), for which all integration measures in the right column of \Tab{MeasureTable} will give $\phi=0$.

Note that before the elements were permuted (\Fig{GraphsFig}, top right), this fact that $\phi=0$ was less obvious.
Moreover, examining all $n!$ permutations (or all $\left({n\atop n/2}\right)$ symmetric bipartitions) would have been an enormously inefficient way of finding that best bipartition for which $\phi$ vanishes. In contrast, finding the connected components of a graph is quite simple, as is evident from staring at \fig{GraphsFig}, with complexity between $O(n)$ and $O(n^2)$.
This means that if we know that $\Phi=0$, then we can find the best bipartition (``cruelest cut") easily, in polynomial time.

Let us now define an approximation taking advantage of this idea: 
{\it replace all unimportant elements $|A_{ij}|<\epsilon$ by zero, and adjust $\epsilon$ so that the largest connected component has size as close as possible to $n/2$.} 
Letting this largest connected component define our approximation of the best bipartition, we now compute its $\phi$-value and use this as our approximation for $\Phi$.

Note that this approximation can be trivially generalized to asymmetric bipartitions 
(the subtle conceptual challenges of how to weight or otherwise handle asymmetric partitions \cite{tononi2008consciousness,balduzzi2008integrated,barrett2011practical,oizumi2014phenomenology} 
are neither ameliorated nor exacerbated by our fast approximation).

In practice, we determine $\epsilon$ by using the interval halving method. A final technical point is that we have two separate definitions of graph connectivity to choose between: weak and strong. A graph is {\it strongly connected} if you can move between any pair of elements following the directional arrows on the edges. This means that every element can (at least through intermediaries) affect and be affected by every other element, precisely capturing the integration spirit of \cite{tononi2008consciousness}. 
Strong connectivity is therefore the logical choice when using our approximation to compute
 $\Phi^{2.5}$, $\Phi^{2.5'}$, $\Phi^{2.5''}$, since it will reflect their property that integration vanishes for afferent and efferent pathways.
 A graph is {\it weakly connected} if you can move between any pair of elements ignoring edge arrows --- in other words, if it simply looks connected when drawn.
Using weak connectivity is arguably the better approximation for the $\Phi$-measures that do not vanish for afferent/efferent pathways, and numerical experiments confirm this.

\begin{figure}[phbt]
\centerline{\includegraphics[width=88mm]{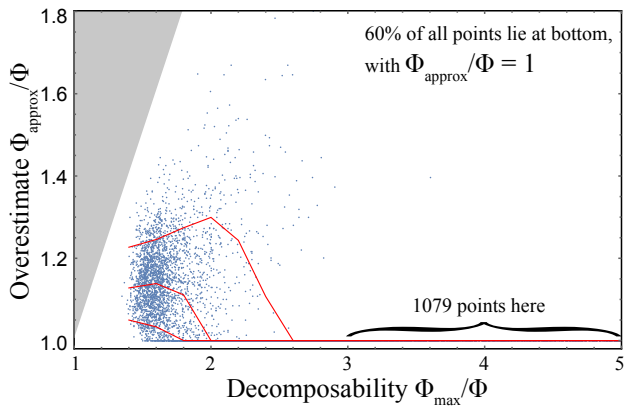}}
\caption{How well our fast $\Phi$-approximation works for 7,000 simulations of the n=16 $\Phi^M$-example described in the text.
Whereas it is seen to be excellent at finding the best bipartition when not all are comparably good,
(\protect\ie, when $\Phi_{\rm max}/\Phi_{\rm min}\gg 1$),  the approximation is seen to overestimate $\Phi$ by up to 15\% (the median) when there is no clear winner (left side). From top to bottom, the three curves show the 95th, 50th and 5th percentiles of the overestimation factor. The shaded region delimits the largest overestimation possible, when $\Phi_{\rm appox}=\Phi_{\rm max}$.
\label{ApproximationFig}
}
\end{figure}

\Fig{ApproximationFig} illustrates the accuracy of our approximation.
For this example,  we randomly\footnote{We generate $\A$-matrices by first computing
\beq{AsimulationEq}
\A=
\eta\A_0+
\left(
\begin{tabular}{cc}
$\A_1$&$0$\\
$0$&$\A_2$
\end{tabular}
\right),
\eeq
where $\A_0$, $\A_1$ and $\A_2$ are random matrices (whose elements are independent Gaussian random variables with zero mean), each normalized to have their largest eigenvalue equal to unity. We then renormalize $\A$ so that its largest eigenvalue equals 0.99.
The parameter $\eta$ controls the typical level of integration: $\eta=0$ gives $\Phi=0$ since $\A$ is block-diagonal, whereas $\eta\to\infty$ gives maximal integration, with no special cut put in by hand; $\eta$ is randomly chosen to be $0.1$, $0.3$, $0.5$, $0.7$, $1$, $2$ or $10$ with equal probability. Once we have generated $\A$, we compute $\C$ as the solution to the Lyapunov equation $\C=\A\C\A^t+\SS$ with $\SS=\I$.
} 
generate 7,000 different $16\times 16$ matrices $\A$ and compute $\Phi^M$ both exactly (as the minimum of $\phi^M$
over all $\left({16\atop 8}\right)=12,870$ symmetric bipartitions) and using our approximation.

For comparison, we also compute the maximum $\Phi^M_{\rm max}$ over the bipartitions.
The ratio $\Phi_{\rm max}/\Phi\ge 1$ (where $\Phi\equiv\Phi_{\rm min}$) 
quantifies how relatively decomposable a system is, whereas the ratio 
$\Phi_{\rm approx}/\Phi\ge 1$ quantifies how well our approximation works, with a value of unity signifying that it is perfect and found the optimal bipartition.
\Fig{ApproximationFig} plots these two quantities against each other, and reveals that they are strongly related. For fairly separable systems, the approximation tends to be excellent: it gives exactly the correct answer 
95\% of the time when $\Phi_{\rm max}/\Phi>2$
and
99.96\% of the time when $\Phi_{\rm max}/\Phi>3$.
When $\Phi_{\rm max}/\Phi\simlt 2$, on the other hand, so that there is less of a clear winner among the bipartitions, our approximation is seen to overestimate the true $\Phi$-value by up to 15\%  on average (this is the median). 

An alternative implementation, which we find works even better for some examples,  is to apply the above-mentioned graph-based bipartition-finding scheme not to the evolution matrix $\A$ but to the covariance matrix $\C$. We therefore recommend computing two approximate bipartitions, one based on $\A$ and one based on $\C$, and selecting the one producing the smaller $\phi$-value.

\section{Conclusions}
\label{ConclusionsSec}

Motivated by the  growing interest in measuring integrated information $\Phi$ in computational and cognitive systems, we have presented a simple taxonomy of $\Phi$-measures where they are each characterized by their choice of factorization method (5 options),  choice of probability distributions to compare ($3\times 4$ options) and choice of measure for comparing probability distributions (5 options). 
We classify all the integration measures revealed in this taxonomy by various desirable properties,  as summarized in \Tab{PropertyTable}.
When requiring the $\Phi$-measures to satisfy a minimum of attractive properties, the hundreds of options reduce to a mere handful, some of which turn out to be identical. All leading contenders are summarized in \Tab{MeasureTable}.

Unfortunately, these most general integration measures are unfeasible to evaluate in practice, with the computational cost growing doubly exponentially with $b$, the number of bits in the system:
they involve a Markov matrix of size $n=2^b$, and they also involve minimizing over approximately 
$N=2^n=2^{2^b}$ bipartitions.
Generalizing the pioneering work of \cite{barrett2011practical}, we derive formulas for the Gaussian case that are exponentially faster, involving manipulations of a matrix whose size grows as $2b$ rather than $2^b$
with the number of variables $b$.
Moreover, we show how the second exponential can also be avoided using an approximation using graph theory, thus reducing the computational cost from doubly exponential to merely polynomial in the system size $b$.

\subsection{Which $\Phi$-measures are best?}

As described in detail in \Sec{BestSec}, six $\Phi$-measures stand out from the taxonomy of hundreds of measures as particularly attractive:
$\Phi^{\rm M}$, 
$\Phi^{\rm M}_{kk'}$,
$\Phi^{3.0}$,
$\Phi^{2.5}$,
$\Phi^{2.5'}$ and
$\Phi^{2.5''}$.
$\Phi^{\rm M}$ retains all the attractive features of the Barrett/Seth measure $\Phi^B$ and adds further improvements: it is guaranteed to vanish for separable systems and to never be negative.
If state-dependence is viewed as desirable, then its cousin $\Phi^{\rm M}_{kk'}$ adds that feature too.

$\Phi^{\rm 3.0}$ is the measure advocated by IIT3.0 and has the many attractive features described in
\cite{oizumi2014phenomenology}. It has the drawback of being the slowest of all the measures to evaluate numerically: its definition involves a linear programming problem which needs to be solved numerically, and even with the fastest algorithms currently available, the computation for a given bipartition grows faster than quadratically with the number of system states --- which in turn grows exponentially with the number of bits, and is infinite for continuous variables.

The remaining three top measures, $\Phi^{2.5}$, $\Phi^{2.5'}$ and $\Phi^{2.5''}$, share with 
$\Phi^{\rm 3.0}$ the arguably desirable feature of vanishing for afferent and efferent systems, but are much quicker to compute.
$\Phi^{\rm 2.5}$ combines core ideas from IIT3.0 with the computational speed of  IIT2.0 \cite{tononi2008consciousness,oizumi2014phenomenology} and elegantly depends only on the system's dynamics and present state, not on any assumptions about which states are more probable. Its drawback of being infinite for continuous variables is overcome by its cousin $\Phi^{\rm 2.5'}$. 

A potential philosophical objection to both $\Phi^{\rm 2.5}$ and $\Phi^{\rm 2.5'}$ is that 
they are arguably not measures of  integration,  but measures of how suboptimal the factorizations $``n"$ and $``m"$ are, since they would both vanish if an optimal factorization were used --- the measure $\Phi^{\rm 2.5''}$ eliminates this concern.

\subsection{Outlook}

Although the results in this paper will hopefully prove useful, there is ample worthwhile work left to do on integration measures.

One major open question is how to best handle asymmetric partitions. We deliberately sidestepped this challenge in the present paper, since it is independent of our results, which is why the subtle normalization issue raised by 
\cite{tononi2008consciousness,balduzzi2008integrated,barrett2011practical,oizumi2014phenomenology} never entered. 
The crux is that if we apply any of the measures in our taxonomy with an asymmetric bipartition, the resulting $\phi$-value will tend to get small when any of the two subsystems is very small,
so simply defining $\Phi$ as the minimum of $\phi$ over all bipartitions (symmetric or not) makes no sense.
IIT3.0 makes an interesting proposal \cite{oizumi2014phenomenology} for how to handle asymmetric partitions, and it is worthwhile exploring whether there are other atttractive options as well. 

Another foundational question is whether our taxonomy can be placed on a firmer logical footing. Although 
our classification based on  factorization, comparison, conditioning and measure may seem sensible and exhaustive, it is interesting to consider whether one or several $\Phi$-measures can be rigorously derived from a small set of attractive axioms alone, in the same spirit as Claude Shannon derived his famous entropy formula,
\eq{SdefEq}.

Yet another foundational question is whether integration maximization can be placed on a firmer physical footing, as advocated by \cite{barrett2014integration,barrett2016comment} in the context of continuous physical fields and by \cite{tegmark2014consciousness} in the context of quantum systems. The formulas in our taxonomy take information, measured in bits, as a starting point. But when I view a brain or computer through my physicist eyes, as myriad moving particles, then what physical properties of the system should be interpreted as logical bits of information?  I interpret as a ``bit" both the position of certain electrons in my computer's RAM memory (determining whether the micro-capacitor is charged) and the position of certain sodium ions in your brain (determining whether a neuron is firing), but on the basis of what principle? Surely there should be some way of identifying consciousness from the particle motions alone, or from the quantum state evolution, even without this information interpretation? If so, what aspects of the behavior of particles corresponds to conscious integrated information? In other words, how can we generalize the quest for neural correlates of consciousness to {\it physical correlates of consciousness}?
IIT argues  that the consciousness occurs at precisely the level of course-graining in space and time that maximizes $\Phi$ \cite{tononi2008consciousness}, which is a prediction that should be tested.

A more practical question involves exploring ways of generalizing and further improving our graph-theory-based approximation for exponential speedup. 
One obvious generalization would involve taking advantage of the structure of $\SS$ (which our method ignored) and the effect of $\x$ (for those $\Phi$-measures that are state-dependent). 
Another interesting opportunity is to generalize from continuous Gaussian systems to arbitrary discrete systems.
For example, if the system consists of $b$ different bits coupled by a nonlinear network of gates, 
one can apply a similar graph-theory approach by defining a $b\times b$ coupling matrix $A_{ij}$ that in some way quantifies how strongly 
flipping the $j^{\rm th}$ bit would affect the $i^{\rm th}$ bit at the next timestep.\footnote{As an example, consider defining 
$A_{ij}$ as the probability that flipping the $j^{\rm th}$ bit will flip the $i^{\rm th}$ bit at the next timestep. If we have six bits evolving according to
$$\x_1=
\left(
\begin{tabular}{c}
$a_1$\\
$b_1$\\
$c_1$\\
$d_1$\\
$e_1$\\
$f_1$
\end{tabular}
\right)
=f(\x_0)=
\left(
\begin{tabular}{c}
$a_0$\\
NOT $a_0$\\
RANDOM\\
$c_0$ XOR $d_0$\\
$c_0$ AND $d_0$\\
$c_0$ AND $d_0$ AND $e_0$
\end{tabular}
\right),
$$
then the coupling matrix is
$$
\A=
\left(
\begin{tabular}{cccccc}
1&0&0&0&0&0\\
1&0&0&0&0&0\\
0&0&0&0&0&0\\
0&0&1&1&0&0\\
0&0&$p_d$&$p_c$&0&0\\
$\>\>$0$\>\>$&$\>\>$0$\>\>$&$p_{de}$&$p_{ce}$&$p_{cd}$&$\>\>$0$\>\>$
\end{tabular}
\right),
$$
were $p_c$ denotes the probability that  $c_0=1$,  $p_{de}$ denotes the probability that $d_0=1$ and $e_0=1$, {\etc}
This coupling matrix is block-diagonal, showing that the bits ${a,b}$ are completely independent of the others.
For a state-independent $\Phi$-measure, these probabilities can be computed as time-averages, otherwise they 
are each zero or one depending on the state. In either case, some elements of the $\A$-matrix can be small but non-zero (making the graph-theory approximation useful)
if the system involves noisy gates or other randomness.
}

As regards practical challenges, it is important to note that there are many other issues besides speed that deserve further work because they have hindered the practical computation of integration $\Phi$-measures from real brain data, including non-stationarity, statistical issues with estimating large numbers of parameters from short data windows without overfitting, possibilities of statistical bias, numerical instabilities, {\etc}

Last but not least, a veritable goldmine of data is becoming available in neuroscience and other fields, and it will be fascinating to measure $\Phi$ for these emerging data sets. In particular, the exponentially faster $\Phi$-measures we have proposed will hopefully facilitate quantitative tests of  theories of consciousness.


\bigskip
\noindent
{\bf Acknowledgements:} 
The author would like to thank Meia Chita-Tegmark, Henry Lin, Adam Barrett, Christof Koch, Masafumi Oizumi and Guilio Tononi for stimulating conversations, useful suggestions and proofreading help
and Dan Fitch for catching typorgaphical errors.
This research was supported by ARO grant W911NF-15-1-0300.

\bibliography{phi}

\end{document}